\documentstyle[12pt]{article}

\newcommand{\be}{\begin{equation}}
\newcommand{\ee}{\end{equation}}
\newcommand{\bea}{\begin{eqnarray}}
\newcommand{\eea}{\end{eqnarray}}

\begin{document}
\begin{titlepage}


\vspace{1in}

\begin{center}
\Large
{\bf THE WORLDSHEET PERSPECTIVE OF T-DUALITY SYMMETRY IN STRING THEORY}

\vspace{.5in}

\normalsize

{ Jnanadeva Maharana \footnote{{\large In Memory of Sumitra Maharana}\\
Invited Review Article for International Journal of Modern Physics A}}  \\ 
E-mail maharana$@$iopb.res.in

\noindent \today

\normalsize
\vspace{.5in}

 {\em Institute of Physics \\
Bhubaneswar - 751005 \\
India  \\
and \\
NISER, Bhubaneswar 751005, India  }

\end{center}

\vspace{.7in}

\baselineskip=24pt

\begin{abstract}
The purpose of this article is to present a pedagogical review of T-duality
in string theory. The evolution of the closed 
string is envisaged on the worldsheet
in the presence of its massless excitations. The duality symmetry is studied
when some of the spacial coordinates are compactified on d-dimensional torus,
$T^d$. The known results are reviewed to elucidate that equations of motion
for the compact coordinates are $O(d,d)$ covariant, $d$ being the number of 
compact directions. Next, the vertex operators of excited massive levels are
considered in a simple compactification scheme. It is shown that
the vertex operators for each massive level can be cast in a T-duality 
invariant form in such a case. Subsequently, the duality properties of 
superstring is investigated in the NSR formulation for the massless backgrounds
such as graviton and antisymmetric tensor. The worldsheet superfield 
formulation is found
to be very suitable for our purpose. The Hassan-Sen compactification is
adopted and it is shown that the worldsheet equations of motion for compact
superfields are $O(d,d)$ covariant when the backgrounds are independent of
superfields along compact directions. The vertex operators for excited levels
are presented in the NS-NS sector and it is shown that they can be cast in 
T-duality invariant form for the case of Hassan-Sen compactification
scheme. An illustrative example is presented to realize our proposal.

\end{abstract}

\vspace{.7in}

\end{titlepage}

\section{Introduction }

\noindent Our principal endeavor in physics is  to comprehend microscopic
laws of Nature from a few fundamental principles. 
Therefore the goal to  unify  the four basic 
 forces is  of paramount importance. The standard model
of particle physics is described by the Electroweak theory which is unification 
of electromagnetic and weak forces and the Quantum Chromodynamics (QCD) is the
underlying theory of the  interaction of quarks and gluons, which is
responsible for strong nuclear force. 
  The 
standard model has been tested to great degree of accuracy.
 There are evidences which
point towards unification of three of the four fundamental interactions i.e.
weak, electromagnetic and strong forces. Moreover, there are compelling
arguments for supersymmetry which might be discovered in accessible high
energy collider experiments. However, experimental discovery of 
supersymmetric partner  particles
is still awaited.  It has been a cherished dream of generations
of physicists to unify the forces of Nature. 
It is recognized that, within the perturbation theoretic 
frame work, the unification of the four forces encounters some difficulties
in the field theoretic approach.  String theory is a radical
step for accomplishing this goal. It 
incorporates gravitational interaction and it 
addresses many
important issues pertaining to quantum gravity. 
The computation of Bekenstein-Hawking
entropy associated with a special class of stringy black holes as derived
from a microscopic theory, like string theory, is one of
the major achievements in this frame work.  
 Furthermore,
the decay rates of special class of stringy  black holes to stringy 
 BPS black holes
have been successfully evaluated and there is no conflict with unitarity
of S-matrix in this context which resolves one of important issues in black 
hole physics. It is also expected that string theory will provide
insights into the conceptual frame works related to the creation of the 
Universe and its
evolution in early epochs. String theory has  provided a basis to explain 
 the mechanism of cosmological inflation. 
Although the standard model of particle physics is so well tested, there
are several reasons to believe that standard model is incomplete to
some extent. For example the Yukawa couplings of the fundamental fermions
in the standard model, which generate their masses, are arbitrary parameters. 
There are several other issues which has stimulated research 
to expound ideas which are beyond the paradigms of the standard model. 
It is hoped that string
theory, being a unified theory of fundamental forces, will address all these
questions collectively. Substantial progress has been made in order
to bring string theory closer to low energy theories of elementary particles.
Thus phenomenological aspects of string theory has attracted considerable
attention. Although the standard model of particle physics, described by
$SU(3)_c \otimes SU(2)\otimes U(1)$, is yet to emerge from string theory,
  interesting
developments are going on to achieve this goal. At this juncture, string theory
is in an interesting phase. There are important developments in the formal
aspects which are bordering and influencing research in 
certain  areas of mathematics. 
On the other hand
there are efforts to establish connections with the phenomenology of
low energy particle physics.\\
The past history as a guide has taught us that symmetry is a guiding principle
to understand physical phenomena and has played crucial role in unraveling
the laws of Nature. For example, the gauge symmetries have been
utilized as cardinal principle to construct the standard model. It is well known
that perturbatively consistent string theories live in critical dimensions:
bosonic strings in 26 dimensions and superstrings in 10 dimensions. Moreover,
in the critical dimensions, there are five perturbatively distinct superstring
theories. As alluded to above, one of the major efforts is to construct
 theories  in four spacetime dimensions in order to establish connections
with the standard model. Therefore, it is proposed that the extra dimensions
(six of them) are very small. In other words the ten dimensional theory is
compactified to four dimensions where the extra spatial 
six dimensions are 
so small that we are unable to probe them using present accessible 
energy scales.
The idea of compactification goes back to Kaluza and Klein. In the
string theoretic scenario, it has very important and interesting implications.
Since string  is 
 a one dimensional object it can wrap around the 'internal'
compactified directions. We do not envisage this possibility in field theoretic
description of a point particle.  Consequently, a rich symmetry structure 
emerges which is a special characteristic of string theory.
 Although there are five different superstring 
 theories in critical dimensions when these theories are compactified to
lower spacetime dimensions through compactification of extra spatial dimensions,
we discover presence of novel symmetries. Some of these symmetries have field 
theoretic analog. We also encounter new types of symmetries which are special
attributes of string theories.The  dualities symmetries, among these,
are very important. The duality symmetries  relate the five string theories 
when we compactify them to lower dimensions. Although the five superstring
theories are perturbatively distinct in the critical dimension, $D=10$,
in the lower spacetime dimensions they get related through the web of 
dualities.
Therefore, it is believed that the five string theories might be different
'phases' of a unique fundamental theory, the M-theory. There are evidences
that such theory might exist, however, a  satisfactory
construction of such a theory is lacking.
\\
The goal of this article is modest and twofold and it is presented in a 
pedagogical manner. The introductory part contains elementary introduction to 
T-duality with some illustrative examples. It is presumed that the reader
is familiar with  the first quantized approach to string theory. It is
expected that the reader has background of the mode expansions, computation
of the energy levels in string theory and is  familiar with results of 
two dimensional
conformal field theory. Moreover, it is 
desirable that reader is acquainted with S-duality and T-duality.  
We shall focus
our attention on the latter from the worldsheet perspective. An interested
reader can use introductory sections of this article to study technical
details in text books and review papers. 
 The second part of this article is a personal point
of view of the author. It has been argued that not only we encounter T-duality
in the massless sector of the closed string but also massive excited levels
of closed string are endowed with the T-duality symmetry. This conjecture is
pursued in the worldsheet approach. More precisely, it is proposed that the
vertex operators associated with the excited massive levels exhibit T-duality
for a string compactified on a torus which will be stated in more detail in
sequel. There are evidences in favor of this conjecture when
vertex operators are studied in a simple compactification scheme. 
Subsequently, the technique
is applied to NSR formalism of superstrings in the presence of NS-NS 
backgrounds. Thus the next two sections (III and IV) contain summary of my
own results. However, I have presented the essential steps for the convenience
of the reader.\\
In recent years a lot of attention has been focused on duality symmetries.
There are excellent text books 
\cite{books,booksa,booksb,booksc,bookd,booke,bookf} 
and  review articles \cite{rev,rev1,rev2,rev3,rev3a,duffm,rev4,reva,revb} 
 on this topic. 
For an
article of this scope and size it is not possible to provide complete references
to all publications. Therefore, keeping in mind the size of the article, I have
provided references to books and review article and some of the articles
published during the nascent period of the field. Any omission is neither
intentional nor it is to deprive any author of due credit.\\
I would like to present my motivations for study of T-duality in a historical
perspective. We recall that the idea of string theory was conceived
from study of collisions of hadrons - strongly interacting particles in
the S-matrix approach. In the low energy regime, the scattering amplitudes
are dominated by nearby resonances and at high energy for small scattering
 angles
they were described by power dependence in energy - the Regge behavior.
The finite energy sum rules (FESR) connected these two seemingly different
attributes of scattering amplitudes \cite{fesr1,fesr2,fesr3}. 
Therefore, the challenge was to
construct an amplitude endowed with both these features, the so called
duality property of that era. The model proposed by Veneziano \cite{venamp}
incorporated those attributes. Indeed construction of
4-point amplitude with the desired features led to the birth of 
string theory. Virasoro \cite{viraamp} 
proposed an amplitude with analogous features. 
Subsequently, multiparticle (n-particle onshell) scattering amplitudes were
constructed. It was soon recognized that Veneziano-type amplitudes 
owe their
origin to tree level open string amplitudes and those of Virasoro are 
derivable from the closed string theory.
As I shall argue in sequel,
this duality symmetry is encoded in the string worldsheet Hamiltonian in its
simplest form. Furthermore, when we envisage evolution of a closed bosonic
string in the presence of 
its massless excitation in some special circumstances the afore
mentioned symmetry appears in a different garb. Therefore, my conjecture is
that, in a very simple scenario, the duality symmetry should be exhibited
in string theory when we consider the scattering of massive states. I
provide evidences in favor of my conjecture in a restricted picture.\\
The plan of the article is as fellows.  In the next section we recall the
rudimentary results of string theory which we need for our discussions. We
introduce simple result of string theory as an illustrative example. Section II
deals with compactification of a string on a torus. We show how T-duality
acts in this theory. Furthermore, it is demonstrated that perturbative spectrum
of the theory remains invariant under the T-duality. Next it is shown,
in the Lagrangian formulation, how the string equations of motion can be cast
in a duality covariant manner \cite{ms}.
 A brief excursion is made to study T-duality
from the effective action point of view for the sake of completeness.
The third section begins by motivation the reader about importance of massive
excited states of strings although most of the works in string theory are
focused on the massless sector since massive states are of the order of
(generally) Planck mass and therefore, they are 
not so  important in the low energy
regime. Some of the well known results about vertex operators of massive states
are recalled and the consequences of conformal invariance, in this context,
are presented. A simple scheme is introduced to examine T-duality properties
of first excited massive level \cite{maha}, 
keeping in mind the remarks of the
preceding paragraph. Furthermore, an analogy is drawn with vertex
operator of massless level when some of the spatial dimensions of the string
are compactified. It is argued that such a simple proposal is
inadequate to study T-duality for vertex operators of arbitrary massive levels
since the vertex  operators assume 
rather complicated form and have large number of
terms. However, encouraged by initial results for first excited massive state,
a prescription is introduced in order to construct manifestly duality
invariant vertex operators for a string whose d-coordinates are compactified
on a torus $T^d$.   The fourth Section is devoted to study of T-duality symmetry
for NSR string in the NS-NS backgrounds when the theory is compactified on
$T^d$. We present two interesting results. We resort to the woldsheet
superfield formulation to study T-duality.
In the first step, we consider,
NSR string in its massless backgrounds in the NS-NS sector, i.e. graviton and 
antisymmetric tensor fields which are independent of superfields along
compact directions. Until recently,
it was not shown explicitly that worldsheet equations of motion in this case, 
i.e. 
for compact coordinates, can be expressed in T-duality covariant form. However,
 for closed bosonic string the equations of motion can be  expressed in
duality covariant form \cite{ms}  as we shall see later. 
We achieved this objective by studying the equations
of motion in the superfield formalism. With this formulation in place, the
path is paved  to construct vertex operators for massive excited states of the
NSR string in the NS-NS sector. We suitably modify the corresponding
formulation for the closed bosonic string to construct T-duality invariant
vertex operators for NSR string in its NS-NS sector. 
 We also present an explicit example of our result for
type IIB theory, in its NS-NS sector when the theory is compactified on
$AdS_3 \otimes S^3 \otimes T^4$. The last section summarizes our results and
future directions. In this section an attempt is made to establish connection
between worldsheet T-duality and recently formulated double field theory.\\

\bigskip

\section{ T-duality Compactified String: Massless Sector }

\noindent
The rich symmetry contents of string theory have played a cardinal role
in understanding diverse attributes of string theory and have
provided deep incisive insights into its dynamics. The target space duality
(T-duality) is a very special feature of string theory and has attracted
considerable attentions over two decades
\cite{books,booksa,booksb,booksc,bookd,booke,bookf,rev,rev1,rev2, rev3, rev3a,rev4,reva,revb}.
This symmetry owes its existence
to one-dimensional nature of string and we do not encounter the analog of
T-duality in field theoretic description of a point particle. Therefore,
it is natural to explore the symmetry from the perspective of the worldsheet.
The T-duality symmetry, associated with closed bosonic string, has been
investigated from the worldsheet perspective in a general framework almost
two decades ago \cite{ms}. The closed bosonic string was
considered in the presence of
its massless backgrounds when  $d$ of its target space coordinates were
compactified on a torus. The backgrounds along compact directions were allowed
to depend on noncompact string coordinates and were assumed
to be independent of compact coordinates. It was recognized that the worldsheet
equations of motion are conserved currents along compact directions.
 The target space duality
(T-duality) is a very special feature of string theory and has attracted
considerable attentions over two decades
\cite{rev1,rev2,rev3,rev3a,rev4,rev,reva,revb}.
We
introduce dual coordinates corresponding to each compact coordinate
and dual backgrounds to derive
another set of equations of motion. The two sets of equations of motion are
 suitably combined to derived $O(d,d)$ covariant equations of motion \cite{ms}
where $d$ is the
number of compact coordinates.\\
On the other hand when T-duality is analyzed in a general setting its
salient features and powerful applications are exhibited from the view point
of the target space in the effective action approach.
To recapitulate, when we
envisage evolution of a closed string in the background of its massless
excitations and demand (quantum) conformal invariance the backgrounds are
constrained through the $\beta$-function equations which are computed
perturbatively in the worldsheet $\sigma$-model approach. These equations of
motion enable us to introduce the effective action whose variation reproduces
the "equations of motion". Let us  toroidally compactify  the
effective action to lower dimensions and examine symmetries of the reduced
effective action. The reduced action can be cast in a manifestly $O(d,d)$
invariant form following the Scherk-Schwarz \cite{ss}
dimensional reduction scheme
if the theory is compactified on a d-dimensional torus ($T^d$) and the
massless backgrounds are independent of the compact coordinates whereas they
carry spacetime dependence.
The availability of a manifestly $O(d,d)$ invariant reduced 
effective action
has been very useful to explore various aspects of string theory in diverse
directions. The target space duality has attracted a lot of attentions
 from different
perspectives over a long period. We refer to some early papers
\cite{revodd,revodda,revoddb,revoddc,revoddd,revodde,revoddf,revoddg,revoddh,
revoddi,revoddj,revoddk,revoddl,revoddm,revoddn,revoddo,revoddp,revoddq}
and
interested reader may consult reviews for comprehensive list of
papers \cite{rev1,rev2,rev3,rev3a,rev4,rev,reva,revb}.\\
The strong-weak duality is a symmetry which relates weak coupling phase to the
strong coupling phase \cite{fonta,shap,rey,ss1,send}. 
This S-duality might relate weak coupling phase of a
theory to the strong coupling phase of another theory. For example, heterotic
string compactified on $T^4$ is S-dual to type IIA theory compactified on
$K_3$. In certain cases they theory might be selfdual. A familiar example is
${\cal N}=4$ supersymmetric Yang-Mills theory. It is obvious that S-duality
symmetry cannot be tested in the perturbative frame work. S-duality together
with T-duality have proved to be a powerful tool to study string dynamics
in diverse dimensions. We mention {\it en passant} that analog of S-duality
is of importance in statistical mechanics where the temperature plays
the role of coupling constant.\\
\bigskip

\noindent{ \bf 2.1 T-Duality: The Woldsheet Description}

\bigskip

\noindent
In this section we briefly recall some of the known and essential results of
string theory. Our attention is focused on closed string theory. Let us
consider the action of a closed string
\bea
\label{poly}
S= {1\over 2}\int d^2\sigma \gamma^{ab}
\partial _aX^{\hat\mu}\partial _bX_{\hat\mu} \eea
where ${\hat\mu}, {\hat\nu}= 0,1,.....{\hat D}-1$, $\hat D$ being the
number of spacetime
dimensions. The above action is the so called gauge fixed action
in the sense that the two dimensional metric is chosen to be the flat metric
since we can transform any metric to a conformally flat metric
and (classically) the conformal factor does not appear. Here $\gamma^{ab}$ is
the two dimensional Lorentzian signature flat metric. We choose this metric
convention here and everywhere.
The equations of motion is
\be
({{\partial ^2}\over{\partial\tau^2}}-{{\partial}\over{\partial\sigma^2}}) X^{\hat\mu}=0
\ee
For a free closed string, $X^{\hat\mu}$ is periodic in $\sigma$
and the solution can be
decomposed into left and right movers i.e $X^{\hat\mu}=
X^{\hat\mu}_L(\tau +\sigma) +X^{\hat\mu}_R(\tau-\sigma)$.
The left and right
movers are expanded in terms of the oscillators in addition to the zero
modes in each sector. The reparametrization invariance symmetry on the
worldsheet, in the quantum theory, requires that the string lives in twenty six
spacetime dimensions.
\\
The canonical Hamiltonian density is
\be
H_c={1\over 2}( P^2+X'^2)
\ee
$P_{\hat\mu}$ being the canonical momentum and $X'^{\hat\mu}$
is derivative of string
coordinate with respect to $\sigma$. Since $P_{\hat\mu}={\dot X}_{\hat\mu}$,
note that
  the Hamiltonian density remains invariant under $P\leftrightarrow X'$, i.e.
when $\tau\leftrightarrow \sigma$ are interchanged. This is the simplest
example of T-duality for a closed string in flat spacetime geometry.\\
In the first quantized approach to string theory, the evolution of the string
is envisaged in its massless excitations such as graviton, $G$ and
 antisymmetric tensor field, B, for a closed string. The worldsheet action
assumes the form,
\be
\label{world}
L={1\over 2}\int d^2\sigma \bigg(\gamma^{ab}G(X)_{\hat\mu\hat\nu}(X)
\partial_aX^{\hat\mu}
\partial_bX^{\hat\nu}+\epsilon^{ab}B_(X){\hat\mu\hat\nu}\partial_aX^{\hat\mu}
\partial_bX^{\hat\nu}\bigg)
\ee
The above action is a two dimensional $\sigma$-model action. The requirements
of conformal invariance imposes constraints on the backgrounds $G$ and $B$
since the associated $\beta$-function must vanish. These constraints assume
forms of differential equations for $G$ and $B$, the so called equations of
motion.\\
In order to explore the world sheet duality symmetry in the simplest
scenario, let us assume
 the backgrounds $G$ and $B$ to be constant i.e. independent of string
coordinates $X^{\hat\mu}$.
In this case the Hamiltonian density is expressed as
\be
H_c=\frac{1}{2}Z^T M(G,B)Z
\ee
where
\be
Z=\pmatrix{ P \cr X' \cr}
\ee
and we have suppressed the indices.
\be
 M=\pmatrix{ G^{-1} & -G^{-1}B \cr
             BG^{-1} & G-BG^{-1}B \cr}
\ee
is a symmetric $2{\hat D}\times 2{\hat D}$ matrix. $G$ and $B$ are
constant backgrounds.

Note that under interchange
$ P\leftrightarrow X'$, the Hamiltonian density remains invariant if we
simultaneously transform $ M\leftrightarrow  M^{-1} $.
The Hamiltonian density is also invariant under the following global
 $\bf{ O(\hat D, \hat D)} $ transformations:
 The $Z$-vector and  $ M$-matrix transform as
\be
Z\rightarrow \Omega_0 Z,~~ M\rightarrow\Omega_0 M\Omega_0 ^T,~
\eta_0\rightarrow\eta_0,~~\Omega_0\in O(\hat D, \hat D)
\ee
 where  $ \eta_0$  is the  $O(\hat D, \hat D)$ metric.
\be
\eta_0=\pmatrix{ 0 &  1 \cr 1 & 0 \cr}
\ee
where  $\bf 1$ is  $ {\hat D}\times {\hat D} $ unit matrix. $\bf Z$ is
$2\hat D$-dimensional $O({\hat D},{\hat D})$ vector.\\
In presence of constant backgrounds, the woldsheet equations of motion for
string coordinates $\{ X^{\hat \mu} ({\sigma, \tau}) \}$ are set
of conservation laws \cite{duff}
\be
\partial _a{\cal J}^a_{\hat\mu}=0
\ee
as follows from (\ref{world}) where the current is given by
\bea
{\cal J}^a_{\hat\mu}=\gamma^{ab} G_{{\hat\mu}{\hat\nu}}
\partial_b X^{\hat\nu}+
\epsilon^{ab} B_{{\hat\mu}{\hat\nu}}\partial_b X^{\hat\nu}
\eea
Thus
locally, one can express the two dimensional current as:
\be
\gamma^{ab}\partial_{b} X^{\hat\nu}G_{\hat\mu\hat\nu}+\epsilon^{ab}\partial_{b}
X^{\hat\nu}B_{\hat\mu\hat\nu}=\epsilon_{ab}\partial_b{\tilde X}_{\hat{\nu}}
\ee
where
$\{{\tilde X}_{\hat\mu} \}$ are set of dual coordinates. The next step is
to introduce a set of auxiliary fields $U^{\hat\mu}_a$ and then define
a dual action
\be
\label{daction}
{\tilde S}=\int d^2\sigma \bigg({1\over 2}
\bigg[( \gamma^{ab}U^{\hat\mu}_aU^{\hat\nu}_b
G_{\hat\mu\hat\nu}+\epsilon^{ab}U^{\hat\mu}_aU^{\hat\nu}_b
B_{\hat\mu\hat\nu}\bigg]\bigg)
\ee
The associated equation of motion is
\be
\label{deq}
\gamma^{ab}U^{\hat\nu}_b G_{\hat\mu\hat\nu}+\epsilon^{ab}U^{\hat\nu}_b
B_{\hat\mu\hat\nu}-\epsilon^{ab}\partial_b{\tilde X}_{\hat\mu} =0
\ee
If we identify $U^{\hat\mu}_a$ with $\partial_a X^{\hat\mu}$ then we recover
original equations of motion.
Thus we can use (\ref{deq}) to solve for $U^{\hat\mu}_a$ in terms of
$\partial_a{\tilde X}^{\hat\mu}$
\be
\label{solu}
U^{\hat\mu}_a=\bigg(\epsilon_a^b{\cal G}^{\hat\mu\hat\nu}+
\delta_a^b{\cal B}^{\hat\mu\hat\nu}\bigg)\partial_b{\tilde X}_{\hat\nu}
\ee
here we have introduced
\be
\label{dualg}
{\cal G}=(G-BG^{-1}B)^{-1}
\ee
 and
\be
\label{dualb}
{\cal B}=-G^{-1}B(G-BG^{-1}B)^{-1}
\ee
Note that ${\cal G}$ and ${\cal B}$ are symmetric and antisymmetric tensors
satisfying $(G+B)({\cal G}+{\cal B})^{-1}=1$. Once we substitute expression
for $U^{\hat\mu}_a$ in (\ref{daction}), 
the dual action assumes the following form
\be
\label{duala1}
{\tilde S}=\int d^2\sigma
\bigg({1\over 2}\bigg[\gamma^{ab}\partial_a{\tilde X}_{\hat\mu}
\partial_b{\tilde X}_{\hat\nu}{\cal G}^{\hat\mu\hat\nu}+\epsilon^{ab}
\partial_a{\tilde X}_{\hat\mu}\partial_b{\tilde X}_{\hat\nu}
{\cal B}_{\hat\mu\hat\nu}\bigg]\bigg)
\ee
We mention in passing that ${\cal G}$ and ${\cal B}$ are also constant dual
backgrounds. Moreover, the ${\tilde X}$ equations of motion derived from
$\tilde S$ are also conservation laws.
Thus the two actions $S$ and ${\tilde S}$ give a pair of equations of motion.
The two sets are equivalent in the sense that they describe the evolution of
the same string theory and the actions are dual to each other. Since equations
from (\ref{duala1}) are worldsheet current conservation law we can identify
(locally) the current as
\be
\label{dualcurl}
\epsilon^{ab}\partial_bX^{\hat\mu}=\gamma^{ab}\partial_b{\tilde X}_{\hat\nu}
{\cal G}^{\hat\mu\hat\nu}+\epsilon^{ab}\partial_b{\cal B}^{\hat\mu\hat\nu}
\ee
In order to examine, how we could expose T-duality from equations of motion,
we can rewrite the equation of motion form $X^{\hat\mu}$ and
${\tilde X}^{\hat\mu}$ in a suitable manner. We multiply the former by
$G^{-1}$ and the latter by ${\cal G}^{-1}$ to arrive at
\be
(G^{-1})^{\hat\mu\hat\nu}\partial_a{\tilde X}_{\hat\nu}-
(G^{-1}B)^{\hat\mu}_{\hat\nu}\partial_a X^{\hat\nu}=\epsilon_a^b\partial_b
X^{\hat\mu}
\ee
and
\be
({\cal G}^{-1})_{\hat\mu\hat\nu}\partial_a X^{\hat\nu}-({\cal G}^{-1}{\cal B})
_{\hat\mu}^{\hat\nu}\partial_a{\tilde X}_{\hat\nu}
=\epsilon_a^b\partial_b{\tilde X}
_{hat\mu}
\ee
Note that these are the two currents whose diverges vanish. Now define
 ${\bar Z}^i=\{X^{\hat\mu}, {\bar X}_{\hat\mu} \}$ which is analogous to
$Z=(P, X')$ which appeared in the definition of the canonical Hamiltonian
density. The above two equations can be combined to a single equation
\be
M\eta\partial_a{\bar Z}=\epsilon_a^b{\bar Z}
\ee
where $M$ is defined in terms of constant backgrounds $G$ and $B$.
Obviously, the current is conserved. Moreover, the equation
is $O({\hat D}, {\hat D})$ covariant. Therefore, we have cast the equations
of motion in a T-duality covariant form once we introduced the dual coordinates
and corresponding dual backgrounds. If the string coordinates are defined on
compactified target manifold, i.e. the coordinates are periodic, then the
dual coordinates will satisfy the same periodicity conditions.
As discussed earlier,
we generally deal with situations where the target manifold is compactified
such that some of the spatial dimensions are quite small and compact. Under
such a circumstance, we are unable to probe them. Therefore, in the low energy
regime, we assume that the background fields (generally massless excitations
of the string) are independent of these coordinates. The simplest
compactification scheme is to assume that the internal space is a torus. The
toroidal compactification might not be the best choice for string theory when
one is attempting to demonstrate emergence of the standard model of particle
physics from string theoretic effective action. However, T-duality plays
a very important role in string theory. Therefore, consequences of
toroidal compactification and underlying symmetries of string theories in this
scheme have been studied very extensively over couple of decades.\\
\bigskip

\noindent{\bf 2.2 Toroidal Compactication and Symmetry of Evolution Equations}

\bigskip
\noindent
We continue with our discussion of T-duality symmetry from the worldsheet
point of view. The next step is to envisage the scenario where we decompose
the string coordinates as follows: $X^{\hat\mu}= (X^{\mu}, Y^{\alpha})$. Here
$X^{\mu}, \mu=0, 1,2...D-1$  are the spacetime coordinates and $Y^{\alpha},
\alpha=D, D+1,  {\hat D}-1$ are toroidally compactified coordinates (d of them)
so that
$D+d={\hat D}$. Moreover, the ${\hat D}$-dimensional backgrounds
$G_{\hat\mu\hat\nu}$ and $B_{\hat\mu\hat\nu}$ are independent of $Y^{\alpha}$.
Since the spacetime (now) is $D$-dimensional, all the tensors should transform
according to the transformation rules of this lower dimensional spacetime.
The components of the ${\hat D}$-dimensional tensors lying along compact
directions transform as scalars from the point of view of $D$-dimensional
spacetime. For example, the $\hat D$-dimensional metric decomposes into
a symmetric tensor,  vectors and scalars (moduli) when the theory is
compactified. Scherk and Schwarz \cite{ss}
have laid down a procedure for decomposition
of tensors under general compactification scheme (even when the internal
manifold has nontrivial curvature). In the context of toroidal
compactification, the procedure becomes relatively simple since torus is
flat. Generally, the dimensional reduction scheme is adopted for string
effective actions; however, in the context of T-duality for strings in the
worldsheet, the tensors are decomposed according to the same prescription.
This aspect is not frequently elaborated in literature. Now we make an
excursion to this topic.
\\
It is most appropriate to adopt the vielbein formalism for the metric
for this purpose \cite{ss}
\bea
\label{schwarz}
 e^{\hat r}_{\hat\mu}=\pmatrix{ e^r_{\mu}(X) &
A^{(1)\beta}_{\mu}(X)E^a_{\beta}(X)\cr 0 & E^a_{\alpha}(X)\cr }
\eea
The spacetime metric is $g_{\mu\nu}=e^r_{\mu}g^{(0)}_{rs}e^s_{\nu}$ and
the internal metric is $G_{\alpha\beta}=E^a_{\alpha}\delta_{ab}E^b_{\beta}$;
$g^{(0)}_{rs}$ is the D-dimensional flat space Lorentzian signature metric.
$ A^{(1)\beta}_{\mu}$ are gauge fields associated with the d-isometries and
it is assumes that the backgrounds depend on coordinates $X^{\mu}$ and are
independent of $Y^{\alpha}$. Similarly, the antisymmetric tensor background,
depending only on $X^{\mu}$ can be decomposed as
\bea
 B_{{\hat\mu}{\hat\nu}} =\pmatrix{B_{\mu\nu}(X) & B_{\mu\alpha}(X)\cr
B_{\nu\beta}(X) & B_{\alpha\beta}(X) \cr}
\eea
Here we note the presence of gauge fields $ B_{\mu\alpha}$ due to
compactification as expected.
The worldsheet action (\ref{world}) will be decomposed into sum of several
terms once we adopt the compactification; however, all backgrounds i.e.
$g_{\mu\nu}, A^{(1)\alpha}_{\mu}, G_{\alpha\beta}, B_{\mu\nu}, B_{\mu\alpha}$
and $B_{\alpha\beta}$ depend only on spacetime coordinates $X^{\mu}$. In what
follows, we closely adopt the technique of \cite{ms} to study T-duality
symmetry through the evolution equations of string coordinates on 
the worldsheet. Some charifications are desirable
 about the notation adopted for
background fields. Here $g_{\hat\mu\hat\nu}$ is the string frame metric in
$\hat D$-dimensions and $g_{\mu\nu}$ is the metric in $D$-dimensions (defined
above). The moduli are $G_{\alpha\beta}$ and $B_{\alpha\beta}$. Note that for
constant backgrounds, we deliberately chose the notation $G_{\hat\mu\hat\nu}$
and $B_{\hat\mu\nu}$ to define the $M$-matrix. We shall define, in the present
case, the $M$-matrix in terms of $G_{\alpha\beta}$ and $B_{\alpha\beta}$ (see
later).
The world sheet action is

\bea
        S = {1\over 2} \int d^2\sigma \bigg( g_{\hat \mu \hat\nu}
\gamma^{ab} +  B_{\hat\mu \hat\nu} \epsilon^{ab}\bigg)\partial_a
X^{\hat\mu} \partial_b X^{\hat\nu}
\eea
Varying this with respect to $X^{\hat\mu} (\sigma, \tau)$ gives the
classical equation of motion for the string
\bea
&&
{\delta S\over\delta X^{\hat\mu}} = -\Gamma_{\hat\mu \hat\nu
\hat\rho} \partial^a X^{\hat\nu} \partial_a X^{\hat\rho} -
g_{\hat\mu \hat\nu} \partial^a \partial_a X^{\hat\nu}\nonumber\\&&
 + {1\over 2} \epsilon^{ab} (\partial_{\hat\mu}  B_{\hat\nu
\hat\rho} + \partial_{\hat\nu}  B_{\hat\rho \hat\mu} +
\partial_{\hat\rho}  B_{\hat\mu \hat\nu}) \partial_a X^{\hat\nu}
\partial_b X^{\hat \rho} = 0
\eea
where
\bea
  \Gamma_{\hat\mu \hat\nu \hat\rho} = {1\over 2}
\bigg( \partial_{\hat\nu}
g_{\hat\mu \hat\rho} + \partial_{\hat\rho}
g_{\hat\mu \hat\nu} - \partial_{\hat\mu}  g_{\hat\nu \hat\rho}\bigg)
\eea
With the insight from the case of  constant backgrounds where we derived
T-duality covariant equations of motion, it is convenient to analyze the
equations of motion for $Y^{\alpha}$ separately first.
Therefore, we look at the action which depends on $Y$,
\bea
S_Y = \int d^2 \sigma \bigg\{ {1\over 2} \bigg(\gamma^{ab}G_{\alpha \beta}(X)
\partial_a Y^{\alpha}
\partial_b Y^{\beta}  + \epsilon^{ab}B_{\alpha \beta}(X) \partial_a
Y^{\alpha} \partial_b Y^{\beta} \bigg) + {\Gamma}^a_{\alpha}(X)
\partial_a Y^{\alpha}\bigg\}
\eea
and
\bea
{\Gamma}^a_{\alpha} = \gamma^{ab} G_{\alpha\beta}
A^{(1)\beta}_{\mu}\partial_b X^{\mu} - \epsilon^{ab}
\big( A^{(2)}_{\mu\alpha} - B_{\alpha\beta}
A^{(1)\beta}_{\mu}\big)\partial_b X^{\mu}
\eea
encodes information about the gauge fields $A^{(1)\alpha}_{\mu}$ and
$A^{(2)}_{\mu\alpha}$. This action
generalizes earlier equations, both by including background vector fields and
by allowing $X$ dependence for all the background fields.
Our aim is to study the equations of motion of $Y$ and suitably modify them
so that these equations are manifestly $O(d,d)$ covariant. Thus we have to
introduce dual coordinates and corresponding backgrounds for the case at
hand. Intuitively we can see that when we consider strings in flat backgrounds,
i.e. massless backgrounds are trivial  $Y$ and $\tilde Y$ would
correspond to the sum and difference of left-moving and right-moving
components. In more general settings, the interpretation is not quite so
simple. \\
Since the backgrounds are independent of $Y^{\alpha}$, the
Euler--Lagrange equations take the form
\be
\partial_a \bigg({\delta S \over \delta \partial_a
Y^{\alpha}} \bigg) = 0
\ee
Therefore, locally, we can write
\bea
{\delta S \over \delta \partial_a Y^{\alpha}} = \gamma^{ab}
\partial_b Y^{\beta} G_{\alpha \beta} + \epsilon^{ab} \partial_b Y^{\beta}
B_{\alpha \beta} + {\Gamma}^a_{\alpha} = \epsilon^{ab}
\partial_b \tilde Y_{\alpha}
\eea
where $\tilde Y_{\alpha}$ are the dual coordinates as before.
They will have
the same periodicities as the $Y^{\alpha}$. Introducing auxiliary fields
$U^{\alpha}_a$, let us now define a dual action for the case at hand
\bea
\tilde S = \int d^2\sigma \bigg\{{1 \over 2} \bigg( \gamma^{ab} U^{\alpha}_a
U^{\beta}_b G_{\alpha \beta} + \epsilon^{ab} U^{\alpha}_a
U^{\beta}_b B_{\alpha \beta}\bigg) + \epsilon^{ab} \partial_a \tilde Y_{\alpha}
U^{\alpha}_b + {\Gamma}^a_{\alpha} U^{\alpha}_a\bigg\}
\eea
If we vary this action with respect to  $\tilde Y_{\alpha}$, we get
$\partial_a(\epsilon^{ab} U^{\alpha}_b)=0$. The auxiliary field equation,
now more complicated, becomes
\bea
\eta^{ab} U^{\beta}_b G_{\alpha \beta} + \epsilon^{ab}
U^{\beta}_b B_{\alpha \beta} - \epsilon^{ab} \partial_b \tilde Y_{\alpha} +
{\Gamma}^a_{\alpha} = 0
\eea
agrees with equation of motion of $Y^{\alpha}$
when we identify $U^{\alpha}_a$ with $\partial_a Y^{\alpha}$.
Although, more complicated compared to constant $G$ and $B$ case, we can solve
for
 for $U^{\alpha}_a$ in terms of $\partial_a
\tilde Y_{\alpha}$ and ${\Gamma}^a_{\alpha}$ and arrive at
\bea
U^{\alpha}_a = \bigg(\epsilon_a{}^b
{\cal G}^{\alpha \beta} +\delta^b_a {\cal B}^{\alpha \beta}\bigg)
\big( \partial_b\tilde Y_{\beta} - \epsilon_{bc} \Gamma^c_{\beta}\big)
\eea
As before
\be
{\cal G} = (G - BG^{-1}B)^{-1}
\ee
and
\be
{\cal B} = - G^{-1}B (G - BG^{-1}B)^{-1}
\ee
Note that (i) the backgrounds depend on string coordinates $X^{\mu}$
and (ii) $(G+B)({\cal G}+{\cal B})=1$, so that ${\cal G}$ and ${\cal
B}$ are the symmetric and antisymmetric parts of $(G+B)^{-1}$,
respectively.
If we substitute for  $U^{\alpha}_a$ in dual action, we arrive at
\bea
\tilde S =&& \int d^2\sigma \bigg\{{1\over 2}\big( \gamma^{ab}
\partial_a \tilde Y_{\alpha} \partial_b \tilde Y_{\beta} {\cal G}^{\alpha
\beta} + \epsilon^{ab} \partial_a \tilde Y_{\alpha}
\partial_b \tilde Y_{\beta} {\cal B}^{\alpha \beta}\big)
- \epsilon^a{}_b \partial_a
\tilde Y_{\alpha} {\Gamma}^b_{\beta} {\cal G}^{\alpha \beta}\nonumber\\&&
- \partial_a \tilde Y_{\alpha} {\Gamma}^a_{\beta} {\cal B}^{\alpha \beta}
- {1 \over 2}\big( \gamma_{ab}
{\Gamma}^a_{\alpha} {\Gamma}^b_{\beta} {\cal G}^{\alpha \beta} +
\epsilon_{ab} {\Gamma}^a_{\alpha} {\Gamma}^b_{\beta} {\cal B}^{\alpha
\beta}\big)\bigg\}
\eea
Following remarks deserve mention (i) ${\cal G}^{\alpha \beta}$ and ${\cal
B}^{\alpha \beta}$ are determined in terms of $G_{\alpha \beta}$ and
$B_{\alpha \beta}$, (ii) they depend only on $X^{\mu}$, (iii) so
does ${\Gamma}^a_{\alpha}$.  Moreover,  the equation of motion derived
from $\tilde S$ is
\be
\partial_a \bigg( {\delta \tilde S \over \delta
\partial_a \tilde Y_{\alpha}} \bigg) = 0
\ee
The two Lagrangians $S$ and $\tilde S$ give a pair of equivalent
equations of motion (at least locally)
\bea
\epsilon^{ab} \partial_b Y^{\alpha} = {\delta \tilde S \over
\delta \partial_a \tilde Y_{\alpha}} = \gamma^{ab}
\partial_b \tilde Y_{\beta} {\cal G}^{\alpha \beta} + \epsilon^{ab}
\partial_b \tilde Y_{\beta} {\cal B}^{\alpha \beta} - \epsilon^a{}_b {\cal
G}^{\alpha \beta} {\Gamma}^b_{\beta} - {\cal B}^{\alpha \beta}
{\Gamma}^a_{\beta}
\eea
In order to express an equation  in an $O(d, d)$
covariant form, we have to combine the pair of equations of motion derived for
$Y^{\alpha}$ and ${\tilde Y}^{\alpha}$ in suitable manner as was done for
constant backgrounds. Although these equations are a lot more complicated,
this goal can be achieved. The pair of equations are
\bea
G^{\alpha \beta} \partial_a \tilde Y_{\beta} -
( G^{-1} B)^{\alpha}{}_{\beta} \partial_a Y^{\beta} = \epsilon_a{}^b
\partial_b Y^{\alpha} + \epsilon_{ab} G^{\alpha \beta}
{\Gamma}^b_{\beta}
\eea
\bea
({\cal G}^{-1})_{\alpha \beta} \partial_a Y^{\beta} - ({\cal G}^{-1}
{\cal B})_{\alpha}{}^{\beta} \partial_a \tilde Y_{\beta} =
\epsilon_a{}^b \partial_b \tilde Y_{\alpha} - \eta_{ab}
{\Gamma}^b_{\alpha} - \epsilon_{ab} ({\cal G}^{-1}
{\cal B})_{\alpha}{}^{\beta}\Gamma^b_{\beta}
\eea
We define the enlarged manifold by combining the compact coordinates and
their corresponding dual coordinates like the previous case. We are
guided by the intuition that the equations of motion for the present case
are still conservation laws (for compact coordinates) although the
worldsheet action for $Y^{\alpha}$ and ${\tilde Y}^{\alpha}$ are lot more
complex.  Defining  $\{ {\tilde Z}^i\} = \{Y^{\alpha} , \, \tilde
Y_{\alpha} \}, \, i = 1, 2, \dots, 2d$, then the above two equations are
combined to a single equation
can be combined as the single equation
\bea
{ M} \eta  \partial_a Z = \epsilon_a{}^b
\partial_b {\tilde Z} + {M} \eta \Sigma_a
\eea
Here $\Sigma_a$ is an $O(d, d)$ vector (for each value of $a$)
given by the column vector
\bea
\Sigma^i_a = \pmatrix {- \gamma_{ab} G^{\alpha \beta} {\Gamma}^b_{\beta} &\cr
\epsilon_{ab} {\Gamma}^b_{\alpha} - \eta_{ab}
B_{\alpha \gamma} G^{\gamma \beta} {\Gamma}^b_{\beta}}
\eea
Note that   $\Sigma_a$ can also re-expressed as
\bea
    \Sigma_a^i = - \partial_a X^\mu {\cal A}_\mu^i + \epsilon_a{}^b
\partial_b X^\mu (M\eta {\cal A}_\mu)^i
\eea
where ${\cal A}_\mu^i$ is comprised of $A_\mu^{(1)\alpha}$ and
$A^{(2)}_{\mu\alpha}$. The former is associated with the $d$-isometries
due to compactification, coming from the metric and the latter are the gauge
fields as we reduce the two form potential to lower dimensions.
Thus we arrive at  the first-order equation
\bea
M\eta (\partial_a Z + {\cal A}_\mu \partial_a X^\mu) =
\epsilon_a{}^b (\partial_b Z + {\cal A}_\mu \partial_b X^\mu)
\eea
This is the desired result. However, we have to still deal with the equations
of motion associated with $X^{\mu}$. Although, these are genuine equations
of motion in the sense that these are not conservation laws due to nontrivial
$X^{\mu}$ dependence carried by all backgrounds, it is important to note that
any transformation carried out along compact directions do not affect the
spacetime tensors and coordinates $X^{\mu}$. Only the moduli and the
gauge field (appearing after dimensional reduction) under go transformations
under T-duality.
However, after some careful manipulations, it can be shown that these
equations are $O(d,d)$ invariant. \\
Now consider the case, when all the string coordinates are
compactified on $\hat D$-dimensional torus, $T^{\hat D}$. We denote all these
coordinates as $Y^{\alpha}$.They satisfy the condition
\be
Y^{\alpha}(\sigma,\tau) +2\pi  = Y^{\alpha}(\sigma,\tau)
\ee
compactification radii is $1$ and the string is still in constant backgrounds.
To distinguish from noncompact coordinates,
we have denoted compact coordinates as $Y^{\alpha}, Y^{\beta}, \alpha, \beta=
0,1, {\hat D}$; this is a special case of compactification scheme
 we have just discussed.
However, we intend to illustrate how the discrete symmetry $O(d,d; Z)$
appears from our perspective.
 Moreover, just for our conveniences, we take spacetime target
space metric to be of Euclidean signature for this particular example.
If we consider compactification of some of the spatial coordinates on $T^d$
such that
\be
X^{\hat\mu}=(X^{\mu}, Y^{\alpha})
\ee
and $Y^{\alpha}$ are compact coordinates then the metric on $T^d$ is
indeed Euclidean.
However, when we leave some coordinates uncompactified, we shall always
consider unusual Lorenzian signature.
Consider motion of a particle on a circular path.
 The momentum is quantized in suitable units of
the inverse radius in order that the wave function is single valued. Next we
consider a massless scalar field, $\phi$ in $\hat D$ dimensions with coordinates
$x^{\hat \mu}$; however, we assume that one of its spatial coordinates
is compact, $S^1$ with radius, $R$. As is well known, the lower
dimensional theory, when $R$ is small has a spectrum consisting of a massless
scalar and a tower of massive states with a spectrum ${n^2}\over {R^2}$.
This is the Kaluza-Klein compactification.
 However, in case of a string,  one of whose coordinate has
geometry of a circle,  offers more interesting possibilities. A distinctive
feature of string theory, with such a compactification is:
we cannot distinguish the perturbative spectrum of this theory
(compactification radius $R$) from that of another string theory whose
coordinate is compactified on a circle of radius $1\over R$. The reason is
that the (closed) string compact coordinate also satisfies periodic boundary
condition and this coordinate can wind around the circle;
\be Y(\sigma,\tau) +2\pi R = Y(\sigma,\tau) \ee
Furthermore, the string coordinate is also periodic when $\sigma$
goes over $2\pi $ for the closed string. Since, the coordinate is
compact, zero momentum mode must be quantized to maintain single
valuedness of the wave function just as the case  in field theory. In
case of the string, the string can wind around the compact direction.
It will cost more energy if the string winds m-number time, because
it will have to stretch more. Therefore, the  effect due to windings has to be
taken into account too  while estimating energy levels.
\be Y_R = y_R +{\sqrt {1\over 2}}p_R(\tau -\sigma) +{\rm oscillators} \ee
\be Y_L = y_L +{\sqrt {1\over 2}}p_L(\tau +\sigma)+{\rm oscillators} \ee
The momentum zero modes $p_{R,L}$  will have the following form to be
consistent with what we said earlier
\be p_R ={{1\over {\sqrt 2}}}({n\over R} -R m), ~~{\rm and} ~~
p_L={{1\over {\sqrt 2}}}({n\over r}+Rm) \ee
Here we have displayed the presence of the compactification radius, $R$, which
is generally set to unity. It is explicitly displayed to demonstrate
$R\rightarrow {1\over R}$ T-duality which interchanges K-K modes and
winding modes.
The above equation states that in general the contribution of the Kaluza-Klein
mode is $1\over R$ times an integer and the winding mode is an integer
times the radius. The total momentum is just $P= {{1\over {\sqrt 2}}}(p_R+p_L),$
which is integral of momentum density over $\sigma$. The total Hamiltonian is
\be H = L_0 +{\bar L}_0 ={1\over 2}(p_L^2 +p_R^2)+{\rm oscillators} \ee
We can generalize the above argument for the case of a closed string with
compact coordinates $Y^{\alpha}(\tau,\sigma)$ in the presence constant massless
backgrounds $G_{\alpha\beta}$ and $B_{\alpha\beta}$.
\\
Now we consider the general case of toroidal compactification and present
the derivation as was done in reference \cite{ms}. Let $G_{\alpha
 \beta}
{\rm and } B_{\alpha \beta}$ be constant backgrounds, $\alpha ,\beta =1,... d$,
and $Y^{\alpha}(\sigma,\tau)$ are the string coordinates. Here $d=\hat D$ since
we take all coordinates to be compact.
 The two-dimensional $\sigma$-model
 The two-dimensional $\sigma$-model
 action containing these coordinates is
\be
S_{compact}={1\over 2} \int d^2 \sigma ~
\bigg[ G_{\alpha\beta} \eta^{ab}\partial_{a}
Y^{\alpha}
\partial_{b} Y^{\beta} + \epsilon^{ab} B_{\alpha\beta} \partial_{a}
Y^{\alpha}
\partial_{b} Y^{\beta} \bigg]\, \ee
where $G_{\alpha\beta}$ and $B_{\alpha\beta}$ are constant backgrounds.
The coordinates are taken to satisfy the periodicity conditions
$Y^{\alpha} \simeq Y^{\alpha} + 2 \pi$. Here we take the compactification
radius to be unity i.e. $R=1$,  for simplicity in calculations.
For closed strings it is necessary that
\be
Y^{\alpha} (2 \pi , \tau) = Y^{\alpha} ( 0, \tau) + 2 \pi m^{\alpha}\,
\ee
where the integers $m^{\alpha}$ are called winding numbers. It follows from
the single-valuedness of the wave function on the torus that the zero modes
of the canonical momentum, $P_{\alpha} = G_{\alpha\beta}
\partial_{\tau} Y^{\beta} + B_{\alpha\beta}
\partial_{\sigma} Y^{\beta}$, are also integers $n_{\alpha}$. Therefore
the zero modes of $Y^{\alpha}$ are given by
\be
Y^{\alpha}_0 = y^{\alpha} + m^{\alpha} \sigma + G^{\alpha\beta}
(n_{\beta} - B_{\beta\gamma} n^{\gamma}) \tau \,\ee
where $G^{\alpha\beta}$ is the inverse of $G_{\alpha\beta}$.
The Hamiltonian is given by
\be
{\cal H} = {1\over 2} G_{\alpha\beta} ( \dot Y^{\alpha} \dot
Y^{\beta} + Y'^{\alpha} Y'^{\beta} )\, \ee
where $\dot Y^{\alpha}$ and $Y'^{\beta}$ are
derivatives with respect to $\tau$ and $\sigma$, respectively. \\
Since $Y^{\alpha} (\sigma , \tau)$ satisfies the free wave equation, we can
decompose it as the sum of left- and right-moving pieces.
The zero mode of $P^{\alpha}=G^{\alpha\beta}P_{\beta}$
is given by $p_L^{\alpha}+p_R^{\alpha}$ where
\be
p^{\alpha}_L = {1\over 2}
[ m^{\alpha} + G^{\alpha\beta} (n_{\beta} - B_{\beta\gamma} m^{\gamma}) ] \ee
\be
 p^{\alpha}_R = {1\over 2}
[ - m^{\alpha} + G^{\alpha\beta} (n_{\beta} - B_{\beta\gamma}
m^{\gamma} ) ]      \ee

The mass-squared operator, which corresponds to the zero mode of ${\cal H}$,
is given (aside from a constant) by

\be
(mass)^2 = G_{\alpha\beta} \big( p^{\alpha}_L p^{\beta}_L + p^{\alpha}_R
p^{\beta}_R \big) +
\sum^{\infty}_{m=1}\sum_{i=1}^d (\alpha^i_{- m} \alpha^{i}_m + \bar
\alpha^{i}_{- m}
\bar\alpha^{i}_m)   \ee
As usual, $\{\alpha_m\}$
and $\{ \bar \alpha_m\}$ denote oscillators associated with
right- and left-moving coordinates,
respectively. Substituting the expressions for $p_L$ and $p_R$,
the mass squared can be rewritten as
\be
\label{zero}
(mass)^2 = {1\over 2} G_{\alpha\beta}
m^{\alpha} m^{\beta} + {1\over 2} G^{\alpha\beta} (n_{\alpha} -
B_{\alpha\gamma} m^{\gamma})(n_{\beta} - B_{\beta\delta} m^{\delta})
+\sum (\alpha^i_{- m} \alpha^i_m + \bar \alpha^i_{- m}
\bar \alpha^i_m) \, \ee
It is significant that the zero mode portion of (\ref{zero}) can be
expressed in the form
      \be  (M_0)^2 = {1\over 2}
(m \ \ n)  M^{-1} \pmatrix {m\cr n\cr}, \ee
where $M$ is the $2d\times2d$ symmetric matrix expressed in terms of constant
backgrounds G and B
\be
\label{mmatrix}
M = \pmatrix {G^{-1} & -G^{-1} B\cr
BG^{-1} & G - BG^{-1} B\cr} \ee
In order to satisfy
$\sigma$-translation symmetry, the contributions of left- and
right-moving sectors to the mass squared
must agree; $L_{0}=\bar L_{0}$.
 The zero mode contribution to
their difference is
\be
\label{diff}
G_{\alpha\beta} (p^{\alpha}_L p^{\beta}_L - p^{\alpha}_R p^{\beta}_R )
= m^{\alpha} n_{\alpha} ~    \ee
Since this is an integer, it always can be compensated by oscillator
contributions, which are also integers. \\
Equation (\ref{diff})  is invariant under interchange of
the winding numbers $m^{\alpha}$ and the discrete momenta $n_{\alpha}$.
Indeed, the
entire spectrum remains invariant if we interchange
$m^{\alpha} \leftrightarrow n_{\alpha}$
 simultaneously let \cite{ms}

\be
(G - B G^{-1} B) \leftrightarrow G^{- 1} ~~~ {\rm
and} ~~~ B G^{- 1} \leftrightarrow - G^{- 1} B \,  \ee
These interchanges precisely correspond to inverting the $2d\times2d$
matrix $M$. This is the spacetime duality
transformation generalizing the well-known duality $R\leftrightarrow
{1\over R}$ in the $d=1$ case discussed earlier.  The general duality
symmetry
implies that the $2d$-dimensional Lorentzian lattice
spanned by the vectors ${\sqrt 2}
(p^{\alpha}_L , \, p^{\alpha}_R)$ with inner product
\be
{\sqrt 2}~ (p_L , \, p_R) \cdot
{\sqrt 2}~ (p'_L , \, p'_R) \equiv 2G_{\alpha\beta} (p^{\alpha}_L p'^{\beta}_L
-
p^{\alpha}_R p'^{\beta}_R) = (m^{\alpha} n'_{\alpha} + m'^{\alpha}
n_{\alpha})\,  \ee
is even and self-dual (\cite{revodde}).

The moduli space parametrized by $G_{\alpha\beta}$ and $B_{\alpha\beta}$
is locally the coset
$O(d, d)/O(d) \times O(d)$.
The global geometry requires also modding out the group of discrete symmetries
generated by $B_{\alpha\beta} \rightarrow B_{\alpha\beta} +
N_{\alpha\beta}$ and $G + B \rightarrow
(G + B)^{-1}$.  These symmetries generate the $O(d,d,Z)$ subgroup of
$O(d,d)$. An $O(d,d,Z)$ transformation is given by a $2d\times2d$ matrix $A$
having integral entries and satisfying $A^T \eta A = \eta$, where $\eta$
consists of off-diagonal unit matrices defined below. Under an $O(d,d,Z)$
transformation
\be \pmatrix {m \cr n} \rightarrow \pmatrix {m' \cr n'} =
A \pmatrix {m \cr n \cr}
\quad {\rm and} \quad M \rightarrow AMA^T\ \ee
It is evident that
\be m\cdot n = {1\over 2}(m \ \ n)
  \eta \pmatrix {m
\cr n \cr}\  \ee
\be \eta=\pmatrix {0 & {\bf 1}\cr {\bf 1} & 0\cr} ,\ee
which appears in eq.(\ref{diff}),
and $M_0^2$ in eq.(\ref{zero})  are preserved under these transformations.
Note that $\eta$ is symmetric $2d\times 2d$ matrix with off diagonal
elements which are d-dimensional unit matrices.
The crucial fact, already evident from the spectrum, is that toroidally
compactified string theory certainly does not share the full $O(d,d)$
symmetry of the low energy effective theory. It is at most invariant
under the discrete $O(d,d,Z)$ subgroup.\\
Let us very briefly discuss the role of T-duality symmetry in open string
theories and we focus on bosonic string theory in order to illustrate
the salient features. When we consider toroidal compactification of a closed
string and examine its spectrum, we discover that both K-K modes and winding
modes contribute to the spectrum besides the excitations due to the action
of oscillators on the vacuum. Since open string has no analog of winding
modes one might think that T-duality has no important roles for open string
theories. However, the open string admits both Dirichlet and Neumann
boundary conditions when we look for solutions to equations of motion. In
recent years, it is recognized that $D_p$ branes play a very important
role in our understanding of string dynamics. We visualize the situation
as follows. These are solitonic objects and they have conformal field theory
descriptions. In $D$-dimensions, if there is a $D_p$-brane, there are
Neumann boundary conditions satisfied in $(p+1)$-diections. These are 
directions of the worldvolume coordinates of $D_p$-brane and we have Dirichlet
boundary conditions along the remaining transverse directions that is
$(D-p-1)$ coordinates satisfy Dirichlet boundary conditions. Thus open
strings can have their end points stuck to these hypersurfances and oscillate.
Put more explicitly, the boundary conditions are as follows
\be
 \partial_{\sigma}X^{\mu}=0, ~{\rm for}~ \mu=0,1,..p-1
\ee
are the Neumann boundary conditions and
\be
X^{\mu}(\sigma=0,\pi)=a^{\mu}_0, ~ {\rm for}~\mu=p,p+1..D-1
\ee
correspond to Dirichlet boundary conditions.
A $D_p$-brane will couple to $p+2$-form RR field strength. Therefore,
$D_0$-brane is interpreted as a particle, 
$D_1$-brane is identified as a D-string and so on. Note that
translational invariance is broken along $\mu=p,..D-1$ and superstring $D=10$.
Recall that T-duality along a give direction can take to a Neumann boundary
condition to Dirichlet or vice versa. Thus a $D_p$-brane can be converted to
a $D_{p+1}$-brane or a $D_{p-1}$-brane as we desire. We elucidated, 
in nutshell, how T-duality can be utilized in the context open strings. 
Open strings oscillate in $d$-dimensions while their end points
are fixed on a $p+1$-dimensional hyperplane and we call it $D_p$ brane.
Thus the open strings whose end points are fixed on these hypersurfaces
satisfy Dirichlet boundary conditions in the $d-p-1$ transverse directions.
However, in those directions, one looses translational invariance. We have
alluded earlier that T-duality along a given direction is equivalent to
$\sigma\leftrightarrow\tau$. Thus, in the context of open string theory,
T-duality operation interchanges Neumann and Dirichlet boundary conditions.
Therefore, in the context of open string theory, when T-duality is implemented
judiciously, one can take a $D_p$ brane to $D_{p+1}$ brane or $D_{p-1}$ brane.
In other words, we are able to realize various types of brane solutions
via T-duality operation as per our requirements.
\\
\bigskip

\noindent{\bf 2.3 Dimensional Reduction of Effective Action and T-duality}

\bigskip

\noindent
So far, in discussing issue compactifications, we have considered situations
when all the coordinates are compact. However, one can envisage the scenario,
when some of the string string coordinates are compactified and the rest are
noncompact. Furthermore, we treated the backgrounds to be constant; however,
in more realistic situations the backgrounds should be allowed to depend on
noncompact coordinates. This is the more interesting situation where we have
a ten dimensional string theory and six of its spatial coordinates are
compactified on a torus $T^6$ so that the resulting theory is reduced to a four
dimensional effective theory. We shall adopt the general prescription of
dimensional reduction \cite{ss,ms,hs}
 so that we can compactify an arbitrary number of
dimensions so that the effective theory is defined in a lower spacetime
dimension, not necessarily four. This will be useful, since the duality
conjectures are in various spacetime dimensions and string
theories are related by the web of dualities in diverse dimensions.\\
The starting point is to consider the string effective action in $\hat D$
spacetime dimensions. The coordinates, metric and all other tensors in the
$\hat D$ dimensional space are specified with a `hat'. The coordinates in
D-dimensional spacetime are denoted by $x^{\mu}, \mu,\nu , etc$ are spacetime
indices.
We discuss a few related points before closing discussions in this section.
When evolution of the closed string is envisaged in the background
of its massless excitations the worldsheet action assumes the form of a
nonlinear $\sigma$-model action. If we impose the constraints of conformal
invariance, these massless backgrounds are required to satisfy
the so called equations of motion since the associated $\beta$-functions
must vanish. If we reconstruct, from these equations of motion, an action
in $\hat D$-dimensional target space such that the Euler-Langrange
equation derived from this action coincides with the equations of motions
obtained from the $\beta$-function equations. Therefore, the
$\hat D$-dimensional 'effective' action is derived order by order in the
$\sigma$-model approach since the $\beta$-function equations are derived
perturbatively from the two dimensional $\sigma$-model. Moreover, generally,
one confines to tree level computation in string perturbation theory in the
sense that the $\beta$-function is computed for the lowest genus Riemann
surface.  We write down the tree level string effective action for the closed
string.
\bea
S_{eff}=\int d^{\hat D}x{\sqrt{-\hat g}}e^{-\hat \Phi}\bigg(R_{\hat g}
+(\partial {\hat \Phi})^2 -{1\over{12}} H_{\hat\mu\hat\nu\hat\rho}
H^{\hat\mu\hat\nu\hat\rho} \bigg)
\eea
where ${ H_{\hat\mu\hat\nu\hat\rho}}=
\partial_{\hat\mu}B_{{\hat\nu}{\hat\rho}}+{\rm cyclic ~ perm} $.
${\hat g}= {\rm det} g_{\hat\mu\hat\nu}$ and
 $ \hat \Phi$ are the 10-dimensional string frame metric and dilaton
respectively. If we are to compactify the
theory, toroidally, to $D$-dimensional spacetime, the  fields appearing
in definition of $R_{\hat g}$ and $ H_{\hat\mu\hat\nu\hat\rho}$ we have define
resulting fields in lower dimensions appropriately. We have already alluded to
the procedure of decomposing the 10-dimensional metric and 2-form fields in
terms of fields in lower dimension. From D-dimensional point of view
these tensors are decomposed to the metric and 2-form, associated gauge fields,
$(A^{(1)\alpha}_{\mu}, A^{(2)}_{\mu\alpha})$ and the moduli, $(G_{\alpha\beta},
B_{\alpha\beta})$. All these fields are independent of the compact coordinates
$y^{\alpha}$.  The next step is to express all the reduced tensors and vectors
in such a way that that their transformations under general coordinate
transformations confirm with those of D-dimensional spacetime. Thus $S_{eff}$
is dimensionally reduced to \cite{ss,ms}
\be
S_{eff}=S_1+S_2+S_3
\ee
where
\be
S_1=\int d^Dx{\sqrt -g}e^{-\phi}\bigg(R_g+(\partial\phi)^2 -
{1\over{12}}H_{\mu\nu\rho}H^{\mu\nu\rho} \bigg)
\ee
\be
 S_2=-{1\over 4}\int d^Dx{\sqrt -g}e^{-\phi}{\cal F}^T_{\mu\nu}
{\cal F}^{\mu\nu}
\ee
\be
 S_3= {1\over 8} \int d^Dx{\sqrt -g}e^{-\phi}\partial_{\mu}M^{-1}
\partial^{\mu}M
\ee
Note that $R_g$ is the scalar curvature computed from the D-dimensional
metric $g_{\mu\nu}$.
$\phi={\hat \Phi}-{1\over 2}{\rm log~det}~G_{\alpha\beta}$ is the shifted
dilaton. $H_{\mu\nu\rho}$ is defined as follows:
\bea
H_{\mu\nu\rho}=\partial_{\mu}B_{\nu\rho}-
{1\over 2}(A^{(1)\alpha}F^{(2)}_{\nu\rho\alpha}+
A^{(2)}_{\mu\alpha}F^{(1)\alpha}_{\nu\rho})+{\rm cyc.perm}
\eea
with $B_{\mu\nu}$ defined as
\bea
B_{\mu\nu}={\hat B}_{\mu\nu}+{1\over 2}A^{(1)\alpha}_{\mu}A^{(2)}_{\nu\alpha}
-{1\over 2}A^{(1)\alpha}_{\nu}A^{(2)}_{\mu\alpha}-
A^{(1)\alpha}_{\mu}B_{alpha\beta}A^{(1)\beta}
\eea
We define ${\hat B}_{\mu\nu}$ to be ${\mu, \nu}$ component of
${\hat D}$-dimensional antisymmetric tensor. The last three terms arise due
to dimensional reduction \cite{ms}. The gauge field strengths
${\cal F}_{\mu\nu}$ are defined as follows which transform as $O(d,d)$ vectors.
and
\bea
{\cal F}_{\mu\nu}=\pmatrix{\partial_{\mu}A^{(1)\alpha}_{\nu}-\partial_{\nu}
A^{(1)\alpha}_{\mu} \cr
\partial_{\mu}A^{(2)}_{\alpha\nu}-\partial_{\nu} A^{(2)}_{\alpha\mu} \cr}
\eea
 and
\bea
M=\pmatrix{ G^{-1} & -G^{-1}B \cr
  BG^{-1} & G-BG^{-1}B \cr}
\eea
is the $M$-matrix defined earlier which depends on spacetime coordinate
$x^{\mu}$. Note that each of the terms $S_1, S_2$ and $S_3$ are $O(d,d)$
invariant on their own. The $O(d,d)$ invariance of the effective action
has played a very important role in generating new solutions from a given
set of backgrounds which follow from equations of motion. In other words,
if we have a set of backgrounds as solutions to $\beta$-function equations
("equations of motion") by implementing $O(d,d)$ transformations judiciously,
we can generate new configurations which are also solutions to equations of
equations. Therefore, we can go from one  string vacuum to another one which
is not connected to the former through any gauge transformation i.e.
general coordinate transformation, gauge transformation of
$A^{(1)\alpha}_{\mu}$ or $A^{(2)}_{\mu\alpha}$ or gauge transformation
associated with two form $B_{\mu\nu}$.  \\
We present a very simple example to illustrate how $O(d,d)$ symmetry is
utilized for generating new solutions starting from a solution of equation of
motion. Consider the cosmological scenario where all backgrounds depend on
the cosmic time $t$. We can write
\bea
g_{\hat\mu\hat\nu}=\pmatrix{1 & 0\cr 0 & G_{ij}\cr}
\eea
\bea
B_{\hat\mu\hat\nu}=\pmatrix{0 & 0\cr 0 & B_{ij} \cr}
\eea
Note that, for this case we can always bring the metric $g_{\hat\mu\hat\nu}$
to this form using a general coordinate transformation and
$G_{ij}, i,j=1,..{\hat D}-1$ is the
spatial part of the metric - can be identifies with $G_{\alpha\beta}$.
Similarly, $B_{\hat\mu\hat\nu}$ also can be brought to the present form using
the gauge transformation on the 2-form B. Its $t-t$ component vanishes from the
antisymmetry property. The shifted dilaton on this occasion is
${\phi}={\hat \Phi} -{1\over 2}{\rm log~ det}G_{ij}$. One can start from a
cosmological solution where $G$ and $\phi$ are nontrivial. Then implement an
appropriate global
$O(D,D)$ transformation (that is the duality group here) and
generate a new cosmological solution where we have ${\tilde G}$, ${\tilde B}$
and $\phi$ such that $G$ and ${\tilde G}$ are not related by general coordinate
transformation. Moreover,  $B$ and ${\tilde B}$ are not connected by the
'vector' gauge transformation of 2-form. The shifted dilaton remains invariant
as is the case with $O(D,D)$ transformation
(see \cite{bookf,matrix2a} for more details).
\section{  Massive Excited States and T-duality }

In this section we explore duality symmetry associated with massive
excited stated of closed string  from the
worldsheet view point. The evolution of the string in the background of its
massless excitation corresponds to a 2-dimensional $\sigma$-model where the
backgrounds are identified as coupling constants of the theory. These are
constrained if we demand that the theory respects conformal
invariance. We intend to follow a similar approach where the string evolves
in the background of higher massive  levels
in order to study the
duality symmetry associated with the excited states.
It is recognized that excited massive stringy states have many interesting
roles in string theory.
\\
Indeed a close examination of string propagation in its massless backgrounds
reveals the importance of excited massive levels in string theory. Let us
follow the arguments put forward by Das and Sathiapalan \cite{bala,bala1}. In order
to investigate implications of conformal invariance, we envisage the
$\sigma$-model in the weak field approximation. To be specific, consider
a closed bosonic string in the graviton background. For sake of simplicity,
when we resort to weak field approximation we write $g_{\mu\nu}=g^0_{\mu\nu}
+h_{\mu\nu}$ where $g^0_{\mu\nu}$ is the flat space Lorentzian metric and
$h_{\mu\nu}$ is the fluctuation (i.e. graviton). Moreover, we expand the
string coordinates as $X^{\mu}(\sigma, \tau)= X^{\mu}_{cl}+\xi^{\mu}$;
$X^{\mu}_{cl}$ being a classical solution of string coordinates and
$\xi^{\mu}$ being the fluctuation. One can compute the $\beta$-function
perturbatively as has been the practice and set it to zero in order to derive
constraints on the background. A covariant formulations is to adopt Riemann
normal coordinate expansion method.
The point made by Das and Sathiapalan is described
in sequel. When one carries out loop expansion for the $\sigma$-model
it is renormalizable at each order in perturbation theory. However, for
some choice of target space, when the loop expansion is summed
to all orders there are new
divergences. These divergences cannot be eliminated from the terms present
in the starting Lagrangian. In order to remove this divergence, it is
essential to introduce higher dimensional operators. In other words,
for the case at hand, the graviton vertex is
$\gamma^{ab}h_{\mu\nu}\partial_a\xi^{\mu}\partial_b\xi^{\nu}$. The new
term has four terms like $\partial \xi\partial \xi\partial \xi\partial \xi$
and contracted with  a fourth rank tensor
(we shall discuss its precise form later).
However, conformal invariance imposes additional constraints on the structure
of the new piece we incorporate i.e. the new vertex operator that is
required to eliminate the fresh divergence.  The resulting constraint
turns out to be precisely the equations of motion for the first excited
massive level of the closed string. Indeed, it describes, in this case
a three point function for graviton-graviton-M, M being (symbolically) the
first excited massive state.  Thus the presence of nonremormalizable
operator of this type has important consequences. The consistency of the
theory will demand, if we include such a term,
 more and more excited levels to be included in the effective
action and eventually the entire tower of stringy states be added.
Of course, string
field theory is the proper arena to address and investigate these issues.
Nevertheless, it is obvious that for the sake of consistency of string
 theory, in the first quantized frame work, excited levels have an important
role.\\
It has been conjectured that excited stringy states might possess local
symmetries \cite{mvx,mv,mva,mvb,ov1,ov1a,ao,kr}.
This idea has been pursued from time
to time and there are evidences that such symmetries manifest themselves,
even in the first quantized approach. A conjecture was first put forward
in the Hamiltonian phase space approach where the local symmetries associated
with graviton and antisymmetric tensor (in the massless sector) were
unraveled through introduction of certain canonical transformations.
In fact, the Ward identities revealed the manifestation of such symmetries
\cite{mvx}.
Subsequently, several authors have carefully studied the proposal and have
found evidences for higher symmetries explicitly for first few massive
levels \cite{ov1,ov1a,ov1b}. Moreover, there have been proposals to explore
possible existence of stringy states at the accessible energy scales
\cite{anto,mas,tom,feng}.\\
The role of excited massive states have come to light in the study of
Planckian energy scattering of string states. For example when
Planckian energy scattering of gravitons are considered, it is essential that
effects of all string states are properly accounted for in order to get
some of the desired features of the scattering amplitude
\cite{planck1,planck2a,planck2b,planck2c,planck2d}.
Moreover, Gross \cite{gross}
has conjectured that in ultra high energy scatterings, when masses of string
states play no significant role, there might be a hidden infinite dimensional
symmetry in string theory. Furthermore, scattering of all stringy state
amplitudes will be related to a single amplitude. Finally, we mention that
Vasiliev's theory of higher spin states has attracted a lot of attentions
in recent years \cite{sagnotti}.
Although the entire programme, with inclusion of
interactions, is yet to be completed, progress has been made to understand
interactions in higher spin field theory from the string theoretic
perspective.\\
In this optics, it is worth while to investigate duality symmetry associated
with excited massive levels of closed string where $d$ of its spacial
coordinates are compactified on $T^d$.
We recall some salient results of T-duality in the frame work of the worldsheet
theory and  we focus on
 toroidal compactification for massless states in the worldsheet approach.\\

\bigskip

\noindent{\bf 3.1 Review of Properties Excited States}

\bigskip

\noindent
Let us very quickly recapitulate some of the results of the previous
section. We shall need these ingredients in what follows.
$Y^{\alpha}(\sigma ,\tau),\alpha , \beta=1,2,..d$ are toroidally
compact coordinate on  $T^d$.
The noncompact coordinates are
$X^{\mu}(\sigma,\tau), \mu ,\nu=0,1,2..D-1$ with $D+d={\hat D}$. The
corresponding backgrounds after dimensional reduction\cite{ms}, for the
  metric, are
$g_{\mu\nu}(X), A^{(1)}_{\mu\alpha}(X) {\rm and}~G_{\alpha\beta}(X)$.
The the 2-form B-field gives
$B_{\mu\nu}(X), B_{\mu\alpha}$ and $B_{\alpha\beta}(X)$ when dimensionally
reduced.
It is assumed that all the backgrounds depend only spacetime
string coordinates
$X^{\mu}$. The gauge fields $A^{(1)}_{\mu\alpha}$ are
associated with the
isometries and $B_{\mu\alpha}$ are another set of gauge fields coming from
dimensional reductions of the 2-form. It was shown, in the previous
section that after introducing
a set of dual coordinates ${\tilde Y}^{\alpha}$ the combined
worldsheet equations of
motion (of $Y ~{\rm and} ~ {\tilde Y}$) can be cast in a duality
covariant form.
Note  that if one resorts to the  Hamiltonian formulation for a slightly
simplified version of above compactification \cite{mahaodd},
the resulting Hamiltonian is expressed in duality invariant form. Our strategy
will be to utilize the results of Hamiltonian formulation and adopt a
simple compactification \cite{hs}
procedure for the higher levels  and unveil the duality
symmetry for these states. Let us consider toroidal compactification where
we set $G_{\alpha\beta}=\delta _{\alpha\beta} ~{\rm and}~ B_{\alpha\beta}=0$;
in other words the radii of $T^d$ are set to unity as before.
The stress energy momentum
tensors used to compute the conformal weights are
\bea
\label{tensors}
T_{++}={1\over 2}\bigg(g^{(0)}_{\mu\nu}\partial X^{\mu}\partial X^{\nu}+
\delta _{\alpha\beta}\partial Y^{\alpha}\partial Y^{\beta} \bigg)
\eea
and
\bea
\label{tensor2}
T_{--}={1\over 2}\bigg(g^{(0)}_{\mu\nu}{\bar\partial}X^{\mu}{\bar\partial}X^{\nu} +
{\bar\partial}Y^{\alpha}{\bar\partial}Y^{\beta}\bigg)
\eea
where $g^{(0)}_{\mu\nu}={\rm diag}(1,-1,-1..)$ is the flat D-dimensional metric,
$\partial X^{\mu}={\dot X}^{\mu}+X'^{\mu}$,
$\partial Y^{\alpha} ={\dot Y}^{\alpha}+Y'^{\alpha}$,
${\bar\partial} X^{\mu}={\dot X}^{\mu}-X'^{\mu}$ and
${\bar\partial}Y^{\alpha}= {\dot Y}^{\alpha}-Y'^{\alpha}$;
dot and 'prime' stand for derivatives with
respect to $\tau$ and $\sigma$ here and everywhere.
We define vertex functions as follows. The vertex operator of a given level is
a sum of several vertex functions. A vertex operator of a given mass level is
required to satisfy $(1,1)$ condition with respect to $(T_{++},T_{--})$.
Consequently, the vertex functions, in general, are not independent and might
satisfy certain relations as we shall see later. In certain cases, some of
them might be 'gauged away' when we count physical degrees of freedom.
We list below the
vertex functions \cite{ov1,ov1a} corresponding to the first massive level
for the uncompactified, ${\hat D}$-dimensional spacetime. A vertex operator is
sum of many vertex functions as given below for the first excited level. 
 As we go to higher and higher levels,
the number of vertex functions increase.
\bea
\label{vetex1}
{\hat V}^{(1)}_1 = A^{(1)}_{{\hat\mu}{\hat\nu} ,{\hat\mu}'{\hat\nu}'}(X)
\partial X^{\hat\mu}\partial X^{\hat\nu}{\bar\partial}X^{{\hat\mu}'}
{\bar\partial}X^{{\hat \nu}'}
\eea
\bea
\label{vertex2}
{\hat V}^{(2)}_1= A^{(2)}_{{\hat\mu}{\hat\nu},{\hat\mu}'}(X)
\partial X^{\hat\mu}\partial X^{\hat\nu}{\bar\partial}^2X^{{\hat\mu}'},~~~
{\hat V}^{(3)}_1= A^{(3)}_{{\hat\mu},{\hat\mu}'{\hat\nu}'}(X)
\partial ^2 X^{\hat\mu}
{\bar\partial}X^{{\hat\mu}'}{\bar\partial}X^{{\hat \nu}'}
\eea
\bea
\label{vertex3}
{\hat V}^{(4)}_1=A^{(4)}_{{\hat\mu},{\hat\mu}'}(X)\partial ^2 X^{\hat\mu}
{\bar\partial}^2X^{{\hat\mu}'}
\eea
The subscript '1' appearing in ${\hat V}^{(1)}_1$ is indicative of the fact
that these vertex functions correspond to ones for the first excited massive
level.
The vertex operator is
\be
{\hat \Phi}_1= \sum_1^4{\hat V}^{(i)}_1
\ee
Notices that the tensor indices are labeled with unprimed and primed indices.
This convention is adopted to keep track of the operators (or oscillators
in mode expansions of $X^{\hat\mu}$) coming from the right moving sector such
as $\partial X^{\hat\mu}$ and from the left moving sector,
${\bar\partial} X^{{\hat\mu}'}$, or powers of $\partial,~ {\bar\partial}$
acting on $X^{\hat\mu}$. It facilitates our future
computation and will be useful
notation when we dwell on duality symmetry in sequel.
It is a straight forward calculation to obtain the constraints on the vertex
functions $V^{(i)}_1$ (actually  conditions on the $X$-dependent tensors,
$A^{(i)}$) if
they are to be $(1,1)$ primaries with respective to $T_{\pm\pm}$. We follow
the methods of \cite{ov1,ov1a} and summarize the relevant results below. These
will be utilized when we explore the associated of T-duality properties of
these vertex operators for the compactified scenario. Note that each of the
vertex functions, ($V^{(2)}_1-V^{(4)}_1$),  is not $(1,1)$ on its own; however,
$V^{(1)}$ is $(1,1)$ as is easily verified. Second point,
we mention in passing,
is that conformal invariance imposes two types of constraints on these vertex
functions: each one satisfies a mass-shell condition (recall that same is true
for tachyon and all massless vertex operators) and gauge (or transversality)
conditions which is also known for all the massless sectors. These are listed
below
\bea
\label{shell1}
({\hat\nabla }^2-2)A^{(1)}_{{\hat\mu}{\hat\nu} ,{\hat\mu}'{\hat\nu}'}(X)=0,~~
({\hat\nabla} ^2-2) A^{(2)}_{{\hat\mu}{\hat\nu},{\hat\mu}'}(X)=0,
\eea
and
\bea
\label{shell2}
({\hat\nabla} ^2-2) A^{(3)}_{{\hat\mu},{\hat\mu}'{\hat\nu}'}(X)=0,~~
({\hat\nabla} ^2-2)A^{(4)}_{{\hat\mu},{\hat\mu}'}(X)=0
\eea
The ${\hat D}$-dimensional Laplacian, ${\hat\nabla}^2$, is
defined in term of the
flat spacetime metric. The mass levels are in in units of the string scale
which has been set to {\it one} in eqs.(\ref{shell1}) and (\ref{shell2}).
The four vertex functions also are related through following equations
\bea
\label{relation}
A^{(2)}_{{\hat\mu}{\hat\nu} ,{{\hat\mu}'}}=
\partial^{{\hat\nu}'}A^{(1)}_{{\hat\mu}{\hat\nu},{{\hat\mu}'}{{\hat\nu}'}},~~~
A^{(3)}_{{\hat\mu},{{\hat\mu}'}{{\hat\nu}'}}=\partial^{\hat\nu}
A^{(1)}_{{\hat\mu}{\hat\nu} ,{\hat\mu}'{\hat\nu}'},~~~
A^{(4)}_{{\hat\mu},{\hat\mu}'}=\partial^{{\hat\nu}'}\partial^{{\hat\nu}}
A^{(1)}_{{\hat\mu}{\hat\nu} ,{\hat\mu}'{\hat\nu}'}
\eea
Here $\partial^{{\hat\mu}}$ etc. stand for partial derivatives with respect to
spacetime coordinates. Furthermore, besides eqs. (\ref{shell1}),(\ref{shell2})
and eq. (\ref{relation}) there are further constraints, like gauge conditions,
which also follow from
the requirements of that the vertex functions be $(1,1)$ primaries
\cite{ov1,ov1a}
\bea
\label{gauge}
{A^{(1){\hat\mu}}_{\hat\mu}},_{{\hat\mu}'{\hat\nu}'}+2\partial^{\hat\mu}
\partial^{\hat\nu}A^{(1)}_{{\hat\mu}{\hat\nu},{{\hat\mu}'}{{\hat\nu}'}}=0,~~~
{\rm and}~~~{A^{(1)}_{{\hat\mu}{\hat\nu},{{\hat\mu}'} }} ^{{\hat\mu}'}+
2\partial^{{\hat\mu}'}
\partial^{{\hat\nu}'}A^{(1)}_{{\hat\mu}{\hat\nu},{{\hat\mu}'}{{\hat\nu}'}}=0
\eea
The above relations, eq.(\ref{relation}) and eq.(\ref{gauge}), will be useful
for our investigation of the duality in what follows.\\
Let us very briefly recapitulate how the T-duality group $O(d,d)$ plays
an important role in the worldsheet Hamiltonian description of a closed
string compactified on $T^d$. We shall proceed in two steps.

However, our
attention will be on the weak field approximation and therefore, we shall
briefly discuss the $O(d,d)$ invariance of graviton vertex operator
(along compact directions),
$h_{\alpha{\beta}'}\partial Y^{\alpha}{\bar \partial}Y^{\beta'}$; the argument
can be extended for the case of weak $B_{\alpha\beta}$ in an analogous manner.
Note that,
in this approximation, the conjugate momentum $P_{\alpha}=\delta_{\alpha\beta}
{\dot Y}^{\alpha}$ and all the indices are raised and lowered by
$\delta ^{\alpha\beta}~{\rm and}~ \delta_{\alpha\beta}$ respectively. Thus
the vertex operator takes the form
\bea
\label{gravvert}
V_h=h^{\alpha{\beta}'}P_{\alpha}P_{\beta}-
h_{\alpha{\beta}'}Y'^{\alpha}Y'^{{\beta}'}
-h^{\alpha}_{{\beta}'}P_{\alpha}Y'^{{\beta}'}+
h^{{\alpha}'}_{{\beta}}P_{{\alpha}'}Y'^{{\beta}}
\eea
We adopt Hassan-Sen compactification scheme where the metric assumes a
block diagonal form (i.e. the gauge fields associated with the isometries
are set to zero)
\bea
g_{\hat\mu\nu}=\pmatrix{g_{\mu\nu} & 0 \cr 0 & G_{\alpha\beta}\cr}
\eea
and correspondingly define the $O(d,d)$ vector
\bea
{\cal W}=\pmatrix{P_{\alpha} \cr Y'^{\alpha}\cr}
\eea
Note that the four terms in (\ref{gravvert}) can combines to express in an
$O(d,d)$  variant form, once we recognize that the first two terms can be
written in terms of the $O(d,d)$ vector $\cal W$ the product $PY'$ needs some
careful handling; we are not canceling out the last two terms in
({\ref{gravvert}) since we continue to maintain distinctions between primed
and unprimed indices.  We
might express $P_{\alpha}$ and $Y'^{\beta}$ as projected $O(d,d)$ vectors
of $\cal W$ contracted with a suitable a suitable tensor and rewrite
(\ref{gravvert}) in the following form
\bea
V_h=H_{mn}{\cal W}^m{\cal W}^n -K_{m}^n{\cal W}^m{\cal W}_n
\eea
Thus $V_h$, above will be $O(d,d)$ invariant $H$ and $K$ if satisfy following
 transformation properties along with the vectors $\{{\cal W}_m \}$
\bea
\label{oddtrans}
H_{mn}\rightarrow \Omega^{m'}_{n}\Omega^{n'}_{n}H_{m'n'},~ {\cal W}^m
\rightarrow \Omega^{m}_{m'}{\cal W}^{m'},~K_m^n\rightarrow
\Omega_{m}^{m'}\Omega^{n}_{n'}K_{m'}^{n'}
\eea
A comment is in order here. We know that $h_{\alpha\beta}$, in $d$-dimensions
has ${d(d+1)}\over 2$ components (when we do not impose tracelessness condition
on $h$). However, counting the number of components of $H_{mn}$ shows that they
exceed those of $h_{\alpha\beta}$. This is not surprising. When we expressed
the Hamiltonian in T-duality invariant form, we introduced $M$-matrix which
belongs to $O(d,d)$. It has $d^2-d$ components whereas $G+B$ have only $d^2$
components. Thus expressing the canonical Hamiltonian in T-duality invariant
form has cost us these extra components. However, the physical degrees of
freedom are the same. A careful analysis \cite{ms} shows that the T-duality
group is ${O(d,d)}\over{O(d)\otimes O(d)}$. On this occasion, we may argue that
we had to pay a price to construct the duality invariant vertex operator. This
argument holds when we construct duality invariant vertex functions for excited
levels.
Note that the inner product of an $O(d,d)$ vector, $T_m$ with ${\cal W}^m$,
 $ T_m{\cal W}^m$ is to be interpreted as follows: $T_m{\cal W}^m=
T^{\alpha}P_{\alpha}+T_{\alpha}Y'^{\alpha}$. Moreover, all $O(d,d)$
tensor indices, $k,l,m,n,...$, are raised and lowered by the
the metric ${\bf \eta}$, whereas the indices of $P_{\alpha}~ {\rm and}~ Y'^{\alpha}$ are raised and lowered by $\delta ^{\alpha\beta}$ and
$\delta_{\alpha\beta}$ respectively.\\

\bigskip

\noindent{\bf 3.2 T-duality Symmetry of Vertex Operators of Excited States}

\bigskip

\noindent
Let us examine T-duality properties of the first excited massive level where
we adopt a simple compactification scheme.
We focus the attention on $V^{(1)}_1$
as an example. Note that if we follow the toroidal compactification scheme
adopted in \cite{ms} in the context of worldsheet duality, for the case at
hand, the vertex function $A^{(1)}_{{\hat\mu}{\hat\nu} ,{\hat\mu}'{\hat\nu}'}(X)
\partial X^{\hat\mu}\partial X^{\hat\nu}{\bar\partial}X^{{\hat\mu}'}
{\bar\partial}X^{{\hat \nu}'}$ will decompose into following forms: (i)
A tensor $A^{(1)}_{\mu\nu ,\mu '\nu '}$, one which has all Lorentz indices
(ii) another which has three Lorentz indices and one index corresponding to
compact directions, (iii) a tensor with two Lorentz indices and two indices
in compact directions, (iv) another,  which has a single Lorentz index and
three indices in in internal directions and (v) a  tensor with all indices
corresponding to compact directions i.e.
$A^{(1)}_{\alpha\beta ,\alpha '\beta '}$. It is obvious these tensors with be
suitably contracted with $\partial X^{\mu}, {\bar\partial}X^{\mu},
\partial Y^{\alpha}, {\bar\partial}Y^{\alpha}$ with all allowed combinations.
We adopt, to start with,  a compactification scheme where only
$A^{(1)}_{\alpha\beta ,\alpha '\beta '}$ is present and the tensors with mixed
indices are absent. We shall return to more general case later.
We may allow the presence of
$A^{(1)}_{\mu\nu ,\mu '\nu '}$; note however, that its presence is not very
essential for the discuss of T-duality symmetry since the spacetime tensors
and coordinates are assumed  to be inert under the T-duality transformations,
as a consequence this term will be duality invariant on its own right.
This is the line of argument advanced by us recently
\cite{maha}.
Therefore, we shall deal with a single vertex function to discuss T-duality
symmetry as a prelude
\bea
\label{vertexy}
V^{(1)}_1= A^{(1)}_{\alpha\beta ,\alpha '\beta '}(X)\partial Y^{\alpha}
\partial Y^{\beta}{\bar\partial}Y^{\alpha '}{\bar\partial}Y^{\beta '}
\eea
As argued earlier, if we expand the expression for
$V^{(1)}_1$, eq.(\ref{vertexy}),
 out in terms of $P_{\alpha}$ and $Y'^{\alpha}$ we get terms of the following
type contacted with the tensor $ A^{(1)}_{\alpha\beta ,\alpha '\beta '}(X)$;
note that we do not use any symmetry(antisymmetry) properties of this tensor
under $\alpha\leftrightarrow\beta$ and $\alpha '\leftrightarrow\beta'$.
Although we express the vertex function in terms of $Y'$ and $P$, we still
like to retain the memory whether these terms came from left movers or right
movers.
The full expression for the vertex function is  classified into five
types. These are listed below:

\noindent (I) All are $P^{\alpha}$'s (index raised by $\delta ^{\alpha\beta}$):
\\
$A^{(1)}_{\alpha\beta ,\alpha '\beta '}(X)P^{\alpha}P^{\beta}P^{\alpha '}
P^{\beta '}$.

\noindent (II) All are $Y'^{\alpha}$'s:\\
$A^{(1)}_{\alpha\beta ,\alpha '\beta '}(X)Y'^{\alpha}Y'^{\beta}
Y'^{\alpha '}Y'^{\beta '}$.

\noindent (III) The four terms with three $P^{\alpha}$'s are:\\
- $A^{(1)}_{\alpha\beta ,\alpha '\beta '}(X)P^{\alpha}P^{\beta}P^{\alpha '}
Y'^{\beta '}$,~~
- $A^{(1)}_{\alpha\beta ,\alpha '\beta '}(X)P^{\alpha}P^{\beta}
Y'^{\alpha '}P^{\beta '}$,\\
$A^{(1)}_{\alpha\beta ,\alpha '\beta '}(X)P^{\alpha}
Y'^{\beta}P^{\alpha '}P^{\beta '}$,~~
$A^{(1)}_{\alpha\beta ,\alpha '\beta '}(X)Y'^{\alpha}P^{\beta}P^{\alpha '}
P^{\beta '}$

\noindent (IV) The four terms with three $Y'^{\alpha}$'s which will eventually
combine with the terms in (III) when we study T-duality property: \\
-$A^{(1)}_{\alpha\beta ,\alpha '\beta '}(X)Y'^{\alpha}Y'^{\beta}P^{\alpha '}
Y'^{\beta '}$,~~
-$A^{(1)}_{\alpha\beta ,\alpha '\beta '}(X)Y'^{\alpha}Y'^{\beta}Y'^{\alpha '}
P^{\beta '}$,\\
$A^{(1)}_{\alpha\beta ,\alpha '\beta '}(X)P^{\alpha}Y'^{\beta}Y'^{\alpha '}
Y'^{\beta '}$,~~
$A^{(1)}_{\alpha\beta ,\alpha '\beta '}(X)Y'^{\alpha}P^{\beta}Y'^{\alpha '}
Y'^{\beta '}$
\noindent (V) There are six terms, each of which is a product of a pair of
momenta ($P^{\alpha}$ and a pair $Y'^{\alpha}$:\\
$A^{(1)}_{\alpha\beta ,\alpha \beta '}P^{\alpha}P^{\beta}Y'^{\alpha '}
Y'^{\beta '}$, ~~
$A^{(1)}_{\alpha\beta ,\alpha \beta '}Y'^{\alpha}Y'^{\beta}P^{\alpha '}
P^{\beta '}$,\\
$- A^{(1)}_{\alpha\beta ,\alpha \beta '}P^{\alpha}Y'^{\beta}P^{\alpha '}
Y'^{\beta '}$,~~
$-A^{(1)}_{\alpha\beta ,\alpha \beta '}Y'^{\alpha}P^{\beta}Y'^{\alpha '}
P^{\beta '}$, \\
$-A^{(1)}_{\alpha\beta ,\alpha \beta '}Y'^{\alpha}P^{\beta}Y'^{\alpha '}
P^{\beta '}$, ~~
$-A^{(1)}_{\alpha\beta ,\alpha \beta '}Y'^{\alpha}P^{\beta}P^{\alpha '}
Y'^{\beta '}$

\noindent A careful inspection of the above terms leads us to conclude
that class (I) and class (II) have the right structures to form an $O(d,d)$
invariant
term when we identify combinations of $P, Y'$ to compose the
$O(d,d)$ vector ${\cal W}$. Similarly, class (III) (with the
product of thee $P$ and one $Y'$) will combine with the class (IV) which
has opposite number of momenta and $Y'$'s. These two (classes) combine to
give us another $O(d,d)$ invariant piece. Note that $\pmatrix{P \cr Y'\cr}$
can be flipped ($P^{\alpha} ~{\rm and}~ Y'^{\alpha}$
interchanged in the column) by operating
the $\bf \eta$-matrix on the ${\cal W}$ vector. It is just like flipping a
down spin Pauli spinor to up spin state.
Finally, the class (V) is a product of a pair of $Y'$
and a pair of $P$; therefore, this class can be cast in duality invariant form.
In order to observe it more transparently, let us consider  the two terms
together in class (I) and in class (II) we can construct two doublets
\bea
\label{doublets}
\pmatrix{A^{(1)}_{\alpha\beta ,\alpha '\beta '} &
A^{(1)}_{\alpha\beta ,\alpha '\beta '} \cr},
~~\pmatrix{{ P}^{\alpha}P^{\beta}P^{\alpha '}P^{\beta '}\cr
Y'^{\alpha}Y'^{\beta}Y'^{\alpha '}Y'^{\beta '} \cr}
\eea
If we take inner  product it will be  T-duality invariant. This procedure can
 be extended to class (III) and class (IV) pairs; moreover the terms in class
(V) can also be cast in the requisite form. However, this prescription is
not very efficient when we consider higher and higher massive levels where
the vertex operators will have increasing number of terms. Therefore, we
propose the following alternative.\\
We can introduce following types of vertex operators in terms of the
$O(d,d)$ vectors ${\cal W}^n$:
\bea
\label{newvertex}
B^{(1)}_{kl,m'n'}(X){\cal W}^k{\cal W}l^l{\cal W}^{m'}{\cal W}^{n'},
B^{(2)}_{kl,m'n'}(X){\cal W}^k({\bf\eta}{\cal W})^{m'}{\cal W}^{n'},
B^{(3)}_{kl,m'n'}(X){\cal W}^k{\cal W}^l{\cal W}^{m'}({\bf\eta}{\cal W})^{n'}
\eea
 $\bf\eta$'s have been inserted to take into account flipping of
$P$ and $Y'$. Notice that the new vertex operators will be $O(d,d)$
invariant if the tensors $B^{(1)}, B^{(2)} ~{\rm and}~ B^{(3)}$ transform as
follows under $O(d,d)$
\bea
\label{btrans}
B^{(i)}{kl,m'n'}\rightarrow \Omega^p_k\Omega^q_l\Omega^{p'}_{m'}\Omega^{q'}_{n'}
B^{(i)}_{pq,p'q'},~~\Omega\in O(d,d)
\eea
since ${\cal W}^k\rightarrow \Omega_{l}^{k}{\cal W}^l$. We draw attention of
the reader to the following points: (i) The prime and unprimed indices
have been maintained even at this stage to keep track of the fact that
certain momenta and $Y'$ originate from the left moving sector and some
other pairs from right moving sector. (ii) The $O(d,d)$ metric $\bf\eta$
is used to raise and lower indices of the corresponding vectors and tensors.
(iii) There are three tensors $B^{(i)}$, their linear combinations are related
to $A^{(1)}$ once one compares all the terms in (\ref{newvertex}) with
the terms collected in class (I) - class (V). (iv) Note also that we have
not included a term which has two $\bf\eta$'s; one introduced between
a pair of ${\cal W}$ coming from right movers and another between left movers.
Such term amounts to 'double flip' and essentially will be equivalent
to the term $B^{(1)}$ since the two $\bf\eta$'s can raise all the indices of
the first vertex.\\
Let us consider the following three vertex functions in the present
compactification scheme
\bea
\label{v3}
 V^{(2)}_1= A^{(2)}_{\alpha\beta ,\alpha '}(X)
\partial Y^{\alpha}\partial Y^{\beta}{\bar\partial}^2Y^{\alpha '},~~~
 V^{(3)}_1= A^{(3)}_{\alpha , \alpha '\beta'}(X)
\partial ^2 Y^{\alpha}\partial Y^{\alpha '}\partial Y^{\beta '}
\eea
and
\bea
\label{v4}
V^{(4)}_1=A^{(4)}_{\alpha ,\alpha '}(X)\partial ^2Y^{\alpha}{\bar\partial}^2
Y^{\alpha '}
\eea
If we demand that these vertex functions be $(1,1)$ primaries then, it follows
from (\ref{relation}) that these vertex operators should vanish, since all
the tensor indices correspond to compact directions whereas the vertex
functions  depend only on $X^{\mu}$. Thus partial derivatives,
$\partial ^{\alpha}$  acting on $A^{(1)}$
vanishes;  thus $A^{(2)}, A^{(3)} ~{\rm and}~ A^{(4)}$, consequently,
 all vanish since they
are related to derivatives of $A^{(1)}$ from the constraints alluded to
earlier (\ref{relation}). However, one issue is to be borne
in mind, in the context
of the T-duality characteristics of these vertex functions,  is that
one does not impose
constraints of conformal invariance to start with.
Note that $V^{(2)}_1$ and $V^{(3)}_1$ have
double derivative on one sector and two single derivatives on another
sector of the world sheet coordinates from dimensional considerations.
Thus we get $\tau$ and/or $\sigma$ derivatives of $P^{\alpha}$,
$Y'^{\alpha '}$ etc. as expected. Consequently,
it becomes quite difficult to cast
vertex functions in a T-duality invariant form when we have such class of
vertex functions ($A^{(4)}$ has similar property now with two double
derivatives on each sector) if we follow the procedure adopted for $V^{(1)}$.
Moreover, we shall continue to encounter these difficulties as we go to
higher and higher excited states. One of the illustrative examples is to
envisage a vertex function for the second massive level which has only
$\partial $ and ${\bar\partial}$ acting of $Y$'s maintaining the required
dimensionality (we are dealing with compactified case here)
\bea
\label{secondmass}
V^{(1)}_2=C^{(1)}_{\alpha\beta\gamma ,\alpha '\beta '\gamma '}
\partial Y^{\alpha}\partial Y^{\beta} \partial Y^{\gamma}
{\bar\partial}Y^{\alpha '}{\bar\partial}Y^{\beta '}{\bar\partial}Y^{\gamma '}
\eea
We can carry out an analysis similar to the vertex function, $V^{(1)}_1$,
of the first massive level and  classify various pieces which can be
combined to construct $O(d,d)$ invariant terms. Indeed, this check,
although tedious, has been done to obtain T-duality invariant
combinations. However, for the 2nd massive level there are vertex functions
of the form,
$\partial^2Y^{\alpha}\partial Y^{\beta}{\bar\partial}^3Y^{\alpha '}$, which are
contracted with suitable tensor and there are many more terms. It is not
easy to exhibit T-duality properties when there are higher order derivatives
of $\sigma$ and $\tau$ acting on $O(d,d)$ vector composed of $P_{\alpha}$
and $Y'^{\alpha}$.\\

\bigskip

\noindent{\bf 3.3 T-duality of Vertex Operators of Arbitrary Massive State}

\bigskip

\noindent
We propose a procedure to systematically organize various vertex operators
which we describe in what follows. (i) The first observation is that the basic
building blocks of vertex functions are
$\partial Y^{\alpha}=P^{\alpha}+Y'^{\alpha}$ and
${\bar\partial} Y^{\alpha '}=P^{\alpha '}-Y'^{\alpha '}$.
(ii) Each vertex function
at a given level is either string of products of these basic blocks or
these blocks are operated by $\partial$ and ${\bar\partial}$ respectively
so that each vertex function at a given mass level has the desired dimensions.
Note that it is not convenient to deal with $P^{\alpha}$ and $Y'^{\alpha}$
separately in order to study
the T-duality properties (in our approach) and the same is true for the
combination $P\pm Y'$. However, $P^{\alpha}$ and $Y'^{\alpha}$ can be projected
out from the $O(d,d)$ vector, $\cal W$.
Similarly, $\partial$ or $\bar\partial$ acting on
$P\pm Y'$ mixed mixed derivatives, in $\sigma$ and $\tau$,
 which is not convenient to deal with
although these partial derivatives create objects of same conformal dimensions.
Let us first introduce following projection operators for later conveniences
\cite{maha}
\bea
\label{project1}
{\bf P}_{\pm}={1\over 2}({\bf 1}\pm {\tilde {\bf {\sigma _3}}})
\eea
where $\bf 1$ is $2d\times 2d$ unit matrix and
$\tilde{\bf{\sigma _3}}=\pmatrix{ {\bf 1} & 0\cr 0 & -{\bf 1} \cr}$
and the diagonal entries $({\bf 1}, -{\bf1})$ stand
for $d\times d$ unit matrices. It is
easy to check that the projection operators are $O(d,d)$ matrices since each
one of them is. We project out two $O(d,d)$ vectors as follows
\bea
\label{vectorproject}
P={\bf P}_+{\cal W}, ~~~Y'={\bf P}_-{\cal W}
\eea
Therefore,
\bea
\label{py}
P+Y'={1\over 2}\bigg({\bf P}_+{\cal W}+ {\bf\eta}{\bf P}_-{\cal W}\bigg),~~
P-Y'={1\over 2}\bigg({\bf P}_+{\cal W}- {\bf\eta}{\bf P}_-{\cal W}\bigg)
\eea
notice that $\bf\eta$ flips lower component $Y'$ vector to an upper component
one. Thus when we have only products  of $P+Y'$ and $P-Y'$, we can express them
first as products of $O(d,d)$ vector and subsequently contract their
indices with appropriate tensors endowed with $O(d,d)$ indices.
Next we deal with worldsheet partial derivatives $\partial$ and $\bar\partial$
operating on basic building blocks. Let us define \cite{maha}
\bea
\label{der}
\Delta _{\tau}={\bf P}_+\partial _{\tau}, ~~~
\Delta _{\sigma}={\bf P}_+{\partial}_{\sigma}~~~{\rm and}~~
{\Delta}_{\pm}(\tau,\sigma)={1\over 2}(\Delta _{\tau}\pm\Delta_{\sigma})
\eea
Therefore,
\bea
\label{pplusy}
\partial (P+Y')=\Delta_+({\tau,\sigma})\bigg({\bf P}_+
{\cal W}+ {\bf\eta}{\bf P}_-{\cal W}\bigg)
\eea
Thus the above expression is an $O(d,d)$ vector. Similarly, when $\bar\partial$
operates on $P-Y'$, we can express it as
\bea
\label{pminusy}
{\bar\partial}(P-Y')=\Delta _-({\tau,\sigma})
\bigg({\bf P}_+{\cal W}- {\bf\eta}{\bf P}_-{\cal W}\bigg)
\eea
The vertex functions we have considered in eqs.(\ref{vertexy}) and
(\ref{secondmass})
which are expressed as only string of products of $\partial Y^{\alpha}$
and ${\bar{\partial}}Y^{\alpha '}$ can be rewritten in terms of the $O(d,d)$
vectors $\cal W$ and subsequently contracted with suitable $O(d,d)$ tensors.
We remind the reader that, now familiar, $M$-matrix which expresses the
Hamiltonian in $O(d,d)$ invariant form  is also parametrized in
terms of backgrounds $G_{\alpha\beta}$ and $B_{\alpha\beta}$. Let us turn our
attentions to the other three vertex functions appearing in (\ref{v3}) and
(\ref{v4}). The procedure outlined above can be adopted to cast $V^{(2)}_1,
V^{(3)}_1$ and $V^{(4)}_1$ in a straight forward manner using the relations
(\ref{pplusy}) and ({\ref{pminusy}). \\
In order to illustrate the variety of vertex functions that can arise as we
envisage higher excited states we list a few vertex operators from
second massive level \cite{ov2}.
\bea
\label{vtwo}
V^{(2)}_2=C^{(2)}_{\alpha\beta,\alpha '\beta '\gamma '}
{\partial}^2Y^{\alpha}\partial Y^{\beta}{\bar\partial}Y^{\alpha'}
{\bar\partial}Y^{\beta'}{\bar\partial}Y^{\gamma ' } +
C^{(3)}_{\alpha\beta\gamma,\alpha '\beta '}\partial Y^{\alpha}\partial Y^{\beta}
\partial Y^{\gamma}{{\bar\partial}^2}Y^{\alpha'}{\bar\partial}Y^{\beta'}
\eea
These vertex operators have a term of the form ${\partial}^2Y^{\alpha}$ or
${{\bar\partial}^2}Y^{\alpha'}$ and rest of the structure is decided by
dimensional considerations. The next class is the one which have
either a  $\partial ^3 Y$ or  ${\bar\partial} ^3 Y$
\bea
\label{vthree}
V^{(3)}_2=C^{(4)}_{\alpha,\alpha'\beta'\gamma'}{\partial}^3Y^{\alpha}
Y^{\alpha'}
{\bar\partial}Y^{\beta'}{\bar\partial}Y^{\gamma ' } +
C^{(5)}_{\alpha\beta\gamma,\alpha'}\partial Y^{\alpha}\partial Y^{\beta}
\partial Y^{\gamma}{{\bar\partial}^3}Y^{\alpha'}
\eea
Another type of term is
\bea
\label{vfour}
V^{(4)}_2=C^{(6)}_{\alpha\beta,\alpha'\beta'}{\partial}^2Y^{\alpha}
\partial Y^{\beta} {{\bar\partial}^2}Y^{\alpha'}{\bar\partial}Y^{\beta'}
\eea
There are  two terms
\bea
\label{vfive}
V^{(5)}_2=C^{(7)}_{\alpha\,\alpha'\beta'}{\partial}^3Y^{\alpha}
 {{\bar\partial}^2}Y^{\alpha'}{\bar\partial}Y^{\beta'}
+C^{(8)}_{\alpha\beta,\alpha'}{\partial}^2Y^{\alpha}\partial Y^{\beta}
 {{\bar\partial}^3}Y^{\alpha'}
\eea
and the last term is
\bea
\label{vsix}
V^{(6)}_2=C^{(9)}_{\alpha,\alpha'}{\partial}^3Y^{\alpha}
{{\bar\partial}^3}Y^{\alpha'}
\eea
The tensors $C^{(2)}-C^{(9)}$ appearing in eqs.(\ref{vtwo}-\ref{vsix}) are
all functions of $X^{\mu}$, independent of compact coordinates $Y^{\alpha}$,
and constrained by requirements of conformal invariance (not necessarily
nonvanishing in the compactification scheme we envisage). We observe from the
structures of vertex operators $V^{(1)}_2 - V^{(6)}_2$ that, each one with
the combinations of the terms will be $O(d,d)$ invariant when we follow
the prescriptions of introducing projection operators, rewrite the
combinations  $P+Y'$ and $P-Y'$ as $O(d,d)$ vectors and
convert $\partial$ and
$\bar\partial$ to ${\Delta_\pm(\tau,\sigma)}$ to operate on
$P\pm Y'$ (re-expressed in terms of the projected ${\cal W}$'s)
respectively. Let us consider $n^{th}$ excited massive level as an example.
The the dimension of all right movers obtained from products of
$\partial Y$ higher powers of $\partial$ acting on $\partial Y$ should be
$(n+1)$ and same hold good for the left moving sector as well. Consider the
right moving sector of the type ${\bf \Pi}_1^{n+1}{\partial Y^{\alpha_i}}$
and the left moving sector ${\bf \Pi}_1^{n+1}{{\bar\partial} Y^{\alpha'_i}}$
The vertex function is
\bea
\label{vertexn}
V_{\alpha_1,\alpha_2...\alpha_{n+1},\alpha'_1\alpha'_2...\alpha'_{n+1}}(X)
{\Large \Pi}_1^{n+1}{\partial Y^{\alpha_i}}
{\Large \Pi}_1^{n+1}{{\bar\partial} Y^{\alpha'_i}}
\eea
and these products of ${\partial Y^{\alpha_i}}$ and
${{\bar\partial} Y^{\alpha'_i}}$ can be converted to products of $(n+1)$
projected ${\cal W}$ for right movers and $(n+1)$ projected ${\cal W}$ from
left movers.
Let us consider a  for vertex function for a high level state.
A  generic vertex will have a structure
\bea
\label{generic}
{\partial}^pY^{\alpha_i}{\partial^q}Y^{\alpha_j}{\partial^r}
Y^{\alpha_k}..
{\bar\partial}^{p'}Y^{\alpha'_i}{\bar\partial}^{q'}
Y^{\alpha'_j}{{\bar\partial}}^{r'}Y^{\alpha'_k}..,~p+q+r=n+1,~p'+q'+r'=n+1
\eea
The product is an $O(d,d)$ tensor whose rank is decided by the constraints on
sum of $p,q ~{\rm and}~ r$ and $p',q'~{\rm and}~ r'$ since number of
$Y^{\alpha_i}$'s and $Y^{\alpha'_i}$'s appearing in (\ref{generic}) is
determined from those conditions. Thus this tensor will be contracted with an
appropriate  tensor
$T_{\alpha_i\alpha_j\alpha_k ..,\alpha'_i\alpha'_j\alpha_k ..}(X)$ which will
give us to a vertex function. Let us discuss how to express
eq.(\ref{generic}) as a product of $O(d,d)$ vectors using the projection
operators introduced earlier.\\
(i) The first step is to rewrite\\
$\partial^pY=\partial^{p-1}(P+Y'),~~~~~{\bar\partial}^{p'}(P-Y')=
{\bar{\partial}}^{p'-1}(P-Y')$\\
(ii) We arrive at\\
$\partial^{p-1}(P+Y')={\Delta_+}^{p-1}(P+Y'),~~
{\bar{\partial}}^{p'-1}(P-Y')={\Delta_-}^{p'-1}(P-Y')$ \\
from (\ref{pplusy}) and (\ref{pminusy})\\
(iii) Finally, using the projection operators (\ref{py}) we get
\\
${\Delta_+}^{p-1}(P+Y')={\Delta_+}^{p-1}\bigg({\bf P}_+{\cal W}+{\bf\eta}
{\bf P}_-{\cal W}\bigg)$,~~
${\Delta}^{p'-1}(P-Y')={\Delta_-}^{p'-1}\bigg(({\bf P}_+{\cal W}-
{\bf\eta}{\bf P}_-{\cal W}\bigg)$
\\
Thus the products in (\ref{generic}) can be expressed as products of
$O(d,d)$ vectors. We need to contract these indices with suitable $O(d,d)$
tensors which have the following form:
\bea
\label{nlevel}
V_{n+1}={\cal A}_{klm..,k'l'm'..}{\Delta_+}^{p-1}{\cal W}_+^k
{\Delta}^{q-1}{\cal W}_+^l
{\Delta_+}^{r-1}{\cal W}_+^m..{\Delta_-}^{p'-1}{\cal W}_-^{k'}
{\Delta_-}^{p'-1} {\cal W}_-^{l'}
{\Delta_-}^{p'-1} {\cal W}_-^{m'}
\eea
where ${\cal W}_{\pm}=({\bf P}_+{\cal W}\pm{\bf\eta}{\bf P}_-{\cal W})$ with
$p+q+r=n+1$ and $p'+q'+r'=n+1$.
Note that superscripts $\{k,l,m;k',l',m'\}$ appearing on ${\cal W}_{\pm}$
in eq. (\ref{nlevel}) are
the indices of the components of the  projected $O(d,d)$ vectors. Moreover,
${\cal A}_{klm..,k'l'm'..}$ is $X$-dependent $O(d,d)$ tensor.\\
Now we turn our attention in another direction. So far we have dealt with
those vertex functions which have only indices corresponding to compact
directions. Therefore, all the massive excitations are scalars under
$ SO(D-1)$. Let us consider the first excited massive level to illustrate
our strategy which can be generalized to any level. We claim that all
vertex functions, for this level, are $O(d,d)$ invariant. We recall $X^{\mu}$
and tensors with only spacetime  indices (i.e. $\mu,\nu,..$etc.)
a tensor transform trivially under the T-duality for these set of indices.
 Thus
\bea
\label{muindex}
{\tilde V}^{(1)}_1={\tilde A}^{(1)}_{\mu\nu,\mu'\nu'}\partial X^{\mu}
\partial X^{\nu}{\bar\partial} X^{\mu'}{\bar\partial} X^{\nu'}
\eea
is $O(d,d)$ invariant as per above prescription. Similarly, vertex functions:\\
 ${\tilde A}^{(2)}_{\mu,\mu'\nu'}\partial^2 X^{\mu}{\bar\partial} X^{\mu'}
{\bar\partial} X^{\nu'}$, ${\tilde A}^{(3)}_{\mu\nu,\mu'}\partial X^{\mu}
\partial X^{\nu}{\bar\partial}^2 X^{\mu'}$ and
${\tilde A}^{(3)}_{\mu,\mu'}\partial^2 X^{\mu}{\bar\partial}^2 X^{\mu'}$
are also $O(d,d)$ invariant. Let us classify the vertex functions according
to the spacetime and 'internal' indices they carry (with appropriate
contractions of course).\\
(A) Vertex functions which have one Lorentz index and three internal indices:
\\
${\tilde B}^{(1)}_{\mu\alpha,\alpha'\beta'}\partial X^{\mu}\partial Y^{\alpha}
{\bar\partial}Y^{\alpha'}{\bar\partial}Y^{\beta'}+$  other terms by permuting
the indices.\\
(B) Vertex functions which have two Lorentz indices and two internal indices:\\
${\tilde B}^{(2)}_{\mu\beta,\mu'\beta'}\partial X^{\mu}\partial Y^{\beta}
{\bar\partial}X^{\mu'}{\bar\partial}Y^{\beta'}+$ other similar terms.\\
(C) Vertex functions with three Lorentz indices and one internal index:\\
${\tilde B}^{(3)}_{\mu\nu,\mu'\beta'}\partial X^{\mu}\partial X^{\nu}
{\bar\partial}X^{\mu'}{\bar\partial}^2Y^{\beta'}+$ other similar terms.\\
(D) Vertex functions of the type:\\
(i)$ {\tilde B}^{(4)}_{\mu\nu,\alpha'}\partial X^{\mu}\partial X^{\nu}
{\bar\partial}Y^{\beta'}+$  other similar terms.\\
(ii) ${\tilde B}^{(5)}_{\mu\beta,\alpha'}\partial X^{\mu}\partial Y^{\alpha}
{\bar\partial}^2Y^{\beta'}+$ other similar terms.\\
(iii) Vertex functions with second derivatives:\\
${\tilde B}^{(6)}_{\alpha,\mu'}\partial^2 Y^{\alpha}{\bar\partial}^2X^{\mu'}$
and ${\tilde B}^{(7)}_{\mu,\alpha'}\partial^2X^{\mu}{\bar\partial}^2Y^{\alpha'}
$\\
The vertex functions whose Lorentz index/indices are contracted with
$\partial X^{\mu},\partial^2 X^{\mu}$,
${\bar\partial}X^{\mu'},\partial^2X^{\mu}$
will be inert under $O(d,d)$ rotations; however, rest of the indices
correspond to internal indices and those are contracted with
$\partial Y^{\alpha}, {\bar\partial}Y^{\alpha'}, \partial^2 Y^{\alpha},
{\bar\partial}^2Y^{\alpha'}$ and so on.
Moreover, the vertex functions considered above, (A)-(D), do not necessarily
vanish unlike the cases when some vertex function, carrying only internal
indices ($V^{(2)}_1$ - $V^{(4)}_1$), vanished as the consequences of conformal
invariance i.e. that these are $(1,1)$ primaries. This conclusion can be
easily verified from relations eqs. (\ref{relation}) and (\ref{gauge}).
We conclude that  only the worldsheet variables with internal indices,
such as $P\pm Y'$  are relevant
to construct $O(d,d)$ vectors which contract with corresponding indices of the
relevant tensors.
We have laid down a procedures to construct $O(d,d)$
vectors from $\partial Y^{\alpha}, {\bar\partial}Y^{\alpha'},
\partial^2 Y^{\alpha},{\bar\partial}^2Y^{\alpha'}$ and other higher derivative
objects. For example,
${\tilde B}^{(1)}_{\mu\alpha,\alpha'\beta'}\partial X^{\mu}\partial Y^{\alpha}
{\bar\partial}Y^{\alpha'}{\bar\partial}Y^{\beta'}$ has three 'internal'
indices of ${\tilde B}^{(1)}$ contracted with $\partial Y^{\alpha}
{\bar\partial}Y^{\alpha'}{\bar\partial}Y^{\beta'}$ and therefore, this
vertex function will be converted to an $O(d,d)$ invariant vertex function
which has a generic form
\bea
\label{convert}
{\tilde T}^{(1)}_{k,k'l'}{\cal W}^k{\cal W}^{k'}{\cal W}^{l'}
\eea
This argument can be carried forward for all vertex
functions at any massive level of the closed bosonic string. Moreover,
the type of vertex functions discussed in (A)-(D) correspond to massive
particles of various spins which fall into the representations of $SO(D-1)$.
Therefore, we are also  able to conclude that vertex functions for massive
levels of a closed bosonic string can be cast in an $O(d,d)$ invariant form for
every level following the procedure presented here.\\



\section{T-duality of NSR String }



\bigskip

\noindent
We further explore the consequences of our proposal in the context of
superstrings. First of all, let us discuss implications of
$\sigma\leftrightarrow\tau$ T-duality for free superstring coordinates.
Under this interchange, the two dimensional chiral bosonic coordinates
transform as
\be
\partial_{\pm}X_{L,R}^{\hat\mu}(\tau ,\sigma)\rightarrow\pm\partial_{\pm}
X_{L,R}^{\hat\mu}( \tau ,\sigma)
\ee
since we can decompose
$X^{\hat\mu}(\tau,\sigma)=X_L^{\hat\mu}(\tau+\sigma)+
X_R^{\hat\mu}(\tau-\sigma)$
When we introduce their (worldsheet) superpartner chiral Majorana fermions,
they transform as
\be
\psi_L^{\hat\mu}(\tau +\sigma)\rightarrow \psi_L^{\hat\mu}( \tau +\sigma),~~~~
\psi_R^{\hat\mu}(\tau-\sigma) \rightarrow -\psi_R^{\hat\mu}(\tau -\sigma)
\ee
It is obvious that the canonical Hamiltonian associated with free theory
will not be $\sigma\leftrightarrow\tau$ T-duality invariant since the
fermionic part of the Lagrangian is first order (it is Dirac Lagrangian) and
consequently the canonical conjugate of a fermion is fermion itself (modulo
$\gamma$ matrix) unlike the bosonic coordinate where the conjugate momentum
is the $\tau$-derivative. Therefore, revealing invariance of the Hamiltonian
density under $O(d,d)$, T-duality, symmetry or alternatively deriving
$O(d,d)$ covariant equations of motion face difficulties if we adopt
an action expressed in terms of component fields. As we shall discuss later,
it is most efficient to go over to the superfield formalism. In the past,
there were certain obstacles to express the worldsheet evolution equations
in $O(d,d)$ covariant form. We have overcome those difficulties recently
\cite{mahans} and
equations of motion can be cast in T-duality covariant form, analogous to
the bosonic case. We have adopted the NSR formalism and our discussions will be
confined to the NS-NS sector all along.\\
A free superstring ( NSR string) action can be
expressed as sum of actions for set of left moving and right moving bosons
and fermions. Therefore, unlike the closed bosonic string case, we do not
see the $P\leftrightarrow X'$ duality (which is same as
$\sigma \leftrightarrow\tau$ duality) so explicitly in the resulting
Hamiltonian density in the presence of fermionic coordinates.
In fact the more transparent duality symmetry is to study
the transformation properties of left moving and right moving fields under
$\sigma\leftrightarrow \tau$ interchange. The holomorphic fields do not
change sign whereas antiholomorphic ones do. If we introduce constant
backgrounds as in the case of closed bosonic string the analog of noncompact
$O({\hat D}, {\hat D})$ symmetry does not emerge so neatly. The target space
duality for superstrings in the NSR formulation has been studied in the past,
however, we feel that this problem deserves further attention.
\\
Let us recapitulate evolution of NSR string in the background of massless
excitations. One starts with the superworldsheet action in two dimensional
superspace where the components of superfield are the bosonic coordinates,
(NSR) Majorana
fermions and auxiliary fields and the backgrounds are functions of the
superfields. We expand the backgrounds in terms of component fields, eliminate
the auxiliary fields  in order to arrive at NSR superstring action in the
presence of massless backgrounds with component
the fermionic and boson fields only.
If we envisage the case where backgrounds
are independent of some of the coordinates (now backgrounds and their
derivatives depend only on bosonic coordinates), then it is very hard to
arrive at duality invariant/covariant equations as was achieved by Maharana and
Schwarz \cite{ms}. Das and Maharana \cite{dm} considered NSR string action in
superspace and adopted the technique introduced by Maharana and Schwarz
\cite{ms} for the closed bosonic string to get analogous equations of motion.
However, they were unable to arrive at at duality covariant equations of
motion although they obtained interesting results for a special case.
In this case the $Z_2$ duality conditions are recovered
\be
\label{z2duality}
\partial_{\pm}X_{L,R}^{\hat\mu}\rightarrow \pm\partial_{L,R}X^{\hat\mu},
~~~\psi^{\mu}_{L,R}\rightarrow \pm\psi^{\mu}_{L,R}
\ee
Moreover, Siegel \cite{warren} considered superstring in superspace in a
Hamiltonian phase space approach to study dualities. Subsequently, there
have been attempts to reveal duality symmetries on superstring
\cite{rest,resta,restb,restc}.
Thus far worldsheet equations of motion for superstrings, transforming
covariantly under duality transformation,
 have  not been derived in a systematic manner at par
with the results of closed bosonic string.
\\

\bigskip

\noindent{\bf 4.1 Superfield Formulation of NSR String and T-duality}

\bigskip

\noindent
The NSR superstring action in two dimensional superspace is given by
\bea
\label{supaction}
S=-{1\over 2}\int d\sigma d\tau d^2\theta{\overline D}{\hat \Phi}^{\hat \mu}
\bigg( G_{{\hat\mu}{\hat\nu}}({\hat\Phi})-\gamma_5
 B_{{\hat\mu}{\hat\nu}}({\hat\Phi})\bigg)D{\hat\Phi}^{\hat{\nu}}
\eea
We have adopted superorthonormal gauge in arriving at this action. Here
${\hat G}_{\hat{\mu\nu}}({\hat \Phi}) $ and
${\hat B}_{\hat{\mu\nu}}({\hat \Phi})$ are the graviton and and 2-form
backgrounds which depend on the superfield ${\hat\Phi}$. It has
expansion in component fields as
\bea
\label{superfield}
{\hat \Phi}^{\hat\mu}= X^{\hat\mu}+{\bar\theta}\psi^{\hat\mu}+
{\bar\psi}^{\hat\mu}\theta+
{1\over 2}{\bar\theta}\theta  F^{\hat\mu}
\eea
where $ X^{\hat\mu}, \psi^{\hat\mu} ~{\rm and}~  F^{\hat\mu}$
are the bosonic, fermionic and auxiliary fields respectively. The covariant
derivatives in superspace are defined to be
\bea
\label{superder}
D_{\alpha}={{\partial}\over{\partial{\bar\theta_{\alpha}}}}-
i(\gamma^a\theta)_{\alpha}\partial_a,~~
{\overline D}_{\alpha}=-{{\partial}\over{\partial\theta_{\alpha}}}+
i({\bar\theta}\gamma^a)_{\alpha}\partial_a
\eea
where $\partial_a$ stands for worldsheet derivatives ($\sigma$ and $\tau$)
and the convention for $\gamma$ matrices are
\bea
\label{gamma}
\gamma^0=\pmatrix{0 & 1\cr 1 & 0\cr},~
\gamma^1=\pmatrix{0 & -1\cr 1 & 0\cr},~
\gamma_5=\gamma^0\gamma^1=\pmatrix{1 & 0\cr 0 & -1\cr}
\eea
The resulting equations of motion from (\ref{supaction}) are
\bea
\label{eq}
{\overline D}\bigg( G_{{\hat\mu}{\hat\nu}}({\hat\Phi})-\gamma_5
B_{{\hat\mu}{\hat\nu}}({\hat\Phi})\bigg)D{\hat\Phi}^{\hat\nu}=0
\eea
The equations for component fields can be obtained by expanding the backgrounds
in terms of them and utilizing the definitions of superspace derivatives
(\ref{superder}). Let us consider a compactification \cite{hs}
scheme such that target
space is compactified on $T^d$: ${\hat{\cal M}}_{\hat D}=M_D\otimes T^d$.
The metric and 2-form backgrounds are decomposed as
\bea
\label{scheme}
 G_{{\hat\mu}{\hat\nu}}=\pmatrix{g_{\mu\nu}(\phi) & 0\cr 0 &
G_{ij}(\phi)\cr}, ~
 B_{{\hat\mu}{\hat\nu}}=\pmatrix{B_{\mu\nu}(\phi) & 0 \cr
0 &  B_{ij}(\phi) \cr}
\eea
Note that the backgrounds only depend on spacetime superfields, $\phi^{\mu}$.
We decompose the superfields as
\be
{\hat\Phi}^{\hat\mu}=(\phi^{\mu}, W^i)
\ee
where $\mu, \nu=0,1,2..D-1$ and $i,j=1,2,..d$ with ${\hat D}=D+d$. Note that
the two superfields have the expansions
\bea
\label{spacetimef1}
 \phi^{\mu}= X^{\mu}+{\bar\theta}{\psi}^{\mu}+
{\bar\psi}^{\mu}\theta+{1\over 2}{\bar\theta}\theta F^{\mu}
\eea
and
\bea
\label{compactf}
 W^i= Y^i+{\bar\theta}{\chi}^i+
{\bar\chi}^i\theta+{1\over 2}{\bar\theta}\theta F^i
\eea
$\chi^i$ are two dimensional Majorana spinors.
In this compactification scheme, the equations of motion for the superfields
$\phi^{\mu}$ is exactly analogous to (\ref{eq}) where we replace $\hat\Phi$
with $\phi$ and the backgrounds with $g_{\mu\nu}(\phi)$ and $B_{\mu\nu}(\phi)$.
\\
Let us focus attention on the evolution equations and the
dynamics of superfields along compact directions \cite{mahans}.
The action is
\bea
\label{waction}
S=-{1\over 2}\int d\sigma d\tau d^2\theta{\overline D}W^i \bigg(G_{ij}(\phi)-
\gamma_5B_{ij}(\phi) \bigg)DW^j
\eea
The superderivatives are defines in (\ref{superder}) above.
The equations of motion for $\{ W^i \}$ are
\bea
\label{wieq}
{\overline D}\bigg(\bigg[G_{ij}(\phi)-\gamma_5 B_{ij}(\phi)\bigg]DW^j\bigg)=0
\eea
In view of the fact that $G$ and $B$ depend only on $\phi^{\mu}$, we may
introduce a dual free superfield ${\widetilde W}_i$ satisfying following
 equation locally which is consistent with (\ref{wieq})
\bea
\label{dualsf}
\bigg(G_{ij}(\phi)-\gamma_5 B_{ij}(\phi) \bigg)DW^j=D{\widetilde W}_i
\eea
and the dual superfield satisfies the constraint:
${\overline D}D{\widetilde W}_i=0$.
Note that (\ref{dualsf}) is reminiscent of the dual coordinate introduced
for closed string by us in the case of closed closed string under a similar
 scenario \cite{ms}. We can go further and
 scenario \cite{ms}. We can go further and
opt for a first order formalism and consider the Lagrangian density
\bea
{\widetilde{\cal L}}={1\over 2}{\bar\Sigma}^i\bigg(G_{ij}(\phi)-
\gamma_5B_{ij}(\phi)
\bigg)\Sigma ^j -{\bar\Sigma}^i D{\widetilde W}_i
\eea
The ${\bar\Sigma}^i$ variation leads to
\be
\bigg(G_{ij}(\phi)-\gamma_5 B_{ij}(\phi) \bigg)\Sigma ^j=D{\widetilde W}_i
\ee
and ${\widetilde W}_i$ variation implies ${\overline D}{\bar \Sigma}^i=0$.
Therefore,
when $\Sigma ^i=DW^i$ we recover (\ref{wieq}).
\\
We are in a position to introduce a dual Lagrangian density in terms of
the dual superfields, ${\widetilde W}_i$ and a set of
dual backgrounds ${\cal G}^{ij}(\phi)$
and ${\cal B}^{ij}(\phi)$; whereas the former of the two backgrounds is
symmetric in its indices the latter is antisymmetric.
\bea
\label{dualw}
{\cal L}_{\widetilde W}= -{1\over 2}{\overline D}{\widetilde W}_i
\bigg({\cal G}^{ij}(\phi)
-\gamma_5 {\cal B}^{ij}(\phi) \bigg)D{\widetilde W}_j
\eea
where the two dual backgrounds, $({\cal G}, {\cal B})$, are related to the
original background fields, $(G, B)$ through the following equations
\bea
\label{dualbg}
{\cal G}=\bigg(G-BG^{-1}B \bigg)^{-1} ~~{\rm and}~~
{\cal B}=-\bigg(G-BG^{-1}B \bigg)^{-1}BG^{-1}
\eea
Notice that $(G-BG^{-1}B)^{-1}$ is symmetric since $(G-BG^{-1}B)$ is symmetric
and it is easy to check that ${\cal B}$ is antisymmetric.
The equations of motion associated with (\ref{dualw}) is
\bea
\label{wideq}
{\overline D}\bigg(\bigg[{\cal G}(\phi)-\gamma_5{\cal B}(\phi) \bigg]
D{\widetilde W}\bigg)=0
\eea
 Our next step is to
write down a pair of equations relating the superfields and their duals
which will lead us to T-duality covariant equations of motion. This is
facilitated
by inspecting the two sets of equations of motion (\ref{wieq}) and
(\ref{wideq})
 resulting from the original
Lagrangian density and its dual which correspond to two   conservation laws.
After straight forward and a little tedious calculations \cite{mahans}
we arrive at following two
equations
\bea
\label{odd1}
DW^i=\gamma_5(G^{-1}B)^i_jDW^j+{G^{-1}}^{ij}D{\widetilde W}_j
\eea
\bea
\label{odd2}
D{\widetilde W}_i=\gamma_5({\cal G}^{-1}{\cal B})^j_iD{\widetilde W}_j+
{\cal G}^{-1}_{ij}DW^j
\eea
Using (\ref{dualbg}) in (\ref{odd2}) we get two sets of equations relating
$DW^i$ and $D{\tilde W}_i$. Let us define a 2d-dimensional $O(d,d)$ vector
(each one is a superfield) such that
\be
\label{oddvec}
{\bf U}=\pmatrix{W^i\cr {\widetilde W}_i\cr}
\ee
and a matrix
\bea
\label{msmatrix}
{\cal M}=\pmatrix{{\bf 1}G^{-1} & \gamma_5G^{-1}B \cr
-\gamma_5 BG^{-1} & {\bf 1}G-{\bf 1}BG^{-1}B \cr}
\eea
where ${\bf 1}$ is the $2\times 2$ unit matrix and $\gamma_5$ is two
dimensional diagonal matrix defined earlier. The ${\cal M}$ matrix has
properties of the familiar $M$-matrix introduced in dimensional reduction of
closed bosonic string: ${\cal M}\in O(d,d)$ and corresponding metric is $\eta$.
The dimensions are further doubled due to the presence of two component
Majorana
fermions.and is reflected by the appearance of $\bf 1$ and $\gamma_5$ in
the ${\cal M}$-matrix.
The two equations (\ref{odd1}) and (\ref{odd2}) can be combined to
a single matrix equation
\bea
\label{oddeqn}
D{\bf U}={\cal M}\eta{\bf U}
\eea
It follows from the definition of the $O(d,d)$ vector ${\bf U}$ that
${\overline D}D{\bf U}=0$.
It holds by virtue of the fact that the two components of
${\bf U}$ satisfy  ${\overline D}DW^i=0$ and ${\overline D}D{\widetilde W}_i=0$
from our original equations (they are dual superfields of each other).
Therefore, we arrive at an $O(d,d)$ covariant equations of motion for
coordinates along compact directions
\bea
\label{finalodd}
{\overline D}\bigg({\cal M}\eta{\bf U}\bigg)=0
\eea
Thus (\ref{finalodd}) generalizes the closed string $O(d,d)$ covariant
equations of motion to NSR superstring \cite{mahans}.
\\
Let us return to the evolution equation for the superfields corresponding
to noncompact coordinates
\bea
\label{noncompact}
{\overline D}\bigg[\bigg(g_{\mu\nu}(\phi)-
\gamma_5 B_{\mu\nu}(\phi) \bigg)D\phi^{\nu}\bigg]=0
\eea
Notice that due to the dependence of backgrounds on the superfield $\phi$
these are "dynamical" equations unlike the case of compact coordinates which
were identified as conservation laws. More important point to note is
that  these equations are T-duality
invariant since the background tensors and these superfields are inert
under the
action of the noncompact T-duality  symmetry group.
Therefore, we conclude that the
resulting equations of motion for a superstring compactified on $T^d$ can be
cast in an $O(d,d)$ covariant form. In the next section, we shall consider
an illustrative example.

\bigskip

\noindent {\bf 4.2 Type IIB Compactification on}
 ${\bf AdS_3\otimes S^3\otimes T^4}$

\bigskip

\noindent
We envisage a scenario where our results can be concretely realized
\cite{mahans}. In the
presence of NS-NS 3-form flux we can write down a worldsheet action for the
case at hand. Notice that $AdS_3$ and $S^3$ correspond to target space of
constant negative and positive curvatures respectively. Therefore, if we
introduce appropriate NS-NS three form fluxes, we can describe the Lagrangian
in these two sectors as sum of two WZW Lagrangians. The presence of WZ term
renders the theory conformally invariant and has the interpretation of the
background antisymmetric tensor fields. Moreover, for the present scenario the
the associated field strengths are constant. The radii of these two spaces are
to be such that the cosmological constants arising from constant positive
and negative curvatures of $S^3$ and $AdS_3$ correspondingly sum up to zero.
The worldsheet description of
NSR string on $AdS_3\otimes S^3$ can be formulated as WZW model on group
manifolds $SL(2,R)\otimes SU(2)$ as is well known
\cite{a1,a1x,a1a,a2,a3,revwzw}. Thus the full worldsheet
action is
decomposed into sum of three parts: one corresponds to superconformal
theory on $SL(2,R)$ the other being $SU(2)$ and the third part is the one
describing a supersymmetric
$\sigma$-model on $T^4$  as we have discussed
in the previous section.\\
Let us briefly consider bosonic WZW model for $SU(2)$ group whose action is
\bea
\label{su2}
S_B &&={1\over {4\lambda^2}}\int d\sigma d\tau{\rm Tr}
\bigg(\partial_ag^{-1}(\sigma,\tau)\partial^ag(\sigma,\tau)\bigg)\nonumber\\
&&+ {k\over{16\pi}}\int_B{\rm Tr}\bigg(g^{-1}(\sigma,\tau)dg(\sigma,\tau)\wedge
g^{-1}(\sigma,\tau)dg(\sigma,\tau)\wedge g^{-1}(\sigma,\tau)dg(\sigma.\tau)
\bigg)
\eea
where $g\in SU(2)$ and satisfies the constraint $gg^{\dagger}={\bf 1}$,
${\bf 1}$ being the unit matrix.
Note the following features: (i) $\lambda$ and $k$ are dimensionless coupling
constants. For a compact group like $SU(2)$, $k$, the coupling constant
appearing in front of the WZ term is quantized for the consistency of the
quantized theory. (ii) g should be smoothly extended to a 3-dimensional
manifold and its boundary, B, is the worldsheet (actually one should define
complex variables in terms of $(\sigma, \tau)$ and this action in those
variables in the standard manner.) The theory is conformally invariant at the
special point ${\lambda ^2}={{4\pi}\over k}$. \\
The case of (bosonic) string on noncompact $SL(2,R)$ manifold is similar to
$SU(2)$ with some differences. (i) The matrix ${\tilde g}\in SL(2,R)$
satisfies the constraint ${\tilde g}\zeta{\tilde g}^T=\zeta$ to be contrasted
with $g\in SU(2)$ group element. Here $\zeta$ is the $SL(2,R)$ metric with
property $\zeta ^2=-{\bf 1}$ and it can be chosen to be
\be
\zeta=\pmatrix{0 & 1\cr -1 & 0\cr}
\ee
(ii) In this case the coefficient of WZ, $k$ term need not be quantized.
\\
We shall consider the supersymmetric WZW model for $SU(2)$ from now on.
The action is \cite{a1,a1x,a1a,a2,a3,revwzw}
\bea
\label{superwzw}
S={1\over {4\lambda^2}}\int d\sigma d\tau d^2\theta{\bar D}{\bf G}^{\dagger}
D{\bf G}
+{k\over{16\pi}}\int d\sigma d\tau d^2\theta\int^1_0dt
{\bf G}^{\dagger}{{{\bf dG}}\over{dt}} {\bar D}{\bf G}^{\dagger}
\gamma_5 D{\bf G}
\eea
The matrix ${\bf G}$ defined in the
superspace satisfies constraints ${\bf G}{\bf G}^{\dagger}={\bf 1}$.
In order to define the WZ term as an integral over a three dimensional
space one defines extension of the superfield to 3-dimensions so that $t=0$
corresponds to the boundary i.e. at that point the two dimensional superfield
is defined on the worldsheet and two dimensional $\gamma_5$ is defined already.
Several remarks are in order at this point: (i)  We express
${\bf G}\in SU(2)$, the  matrix  in terms of component fields. When
the  auxiliary
field is eliminated from the action and  the $d\theta$ integration is done,
the resulting action
contains quartic fermionic coupling and the theory is not necessarily
conformally invariant for arbitrary $\lambda^2$ and $k$. (ii) At the special
point $\lambda^2={{4\pi}\over k}$, theory is conformally invariant and the
quartic fermionic coupling disappears. Therefore, for a superstring on a group
manifold the two coupling constants are related (and $k$ is quantized).
Moreover, at the conformal point, the equations of motion of the superfields
(${\bf G}$-matrices), decompose into holomorphic and antiholomorphic parts and
the equations of motion
 take the form of two separate current conservation equations. This feature
is most elegantly displayed if we expand the super-matrix in ${\bf G}$
in light cone variables as
\bea
\label{lcexp}
{\bf G}(\sigma,\tau,\theta)
=g(\sigma,\tau)\bigg( {\bf 1}+i\theta^+\psi_+(\sigma,\tau)
+i\theta^-\psi_-(\sigma,\tau)+i\theta^+\theta^-F(\sigma,\theta) \bigg)
\eea
Corresponding light cone superderivatives are
\bea
\label{superlc}
D_{\pm}={{\partial}\over{\partial\theta^{\pm}}}-i\theta^{\pm}\partial_{\pm}
~~{\rm with}~~\partial_{\pm}=\partial_{\tau}\pm\partial_{\sigma}
\eea
Note that the chiral fermions $\psi_{\pm}$ are matrices taking value in the
Lie algebra and $F$ is the auxiliary field. The constraint $GG^{\dagger}=
{\bf 1}$ results in relations between component fields, $g,\psi_{\pm}$ and
$F$. At the conformal point the equation of motions become
\bea
\label{seqn}
D_{\mp}{\cal {\bf J}}_{\pm}=0, ~~~~
{\cal{\bf J }}_{\pm}=-i{\bf G}^{-1}D_{\pm}{\bf G}
\eea
We therefore, note that classically we solve for NSR string on $S^3$.\\
The case of NSR string on $AdS_3$, proceeds similarly once we take into account
the subtleties associated with the noncompact $SL(2,R)$ group.\\
We have discussed in detail how to construct worldsheet action for compact
coordinates in the case of NSR string on $T^d$. We showed that the equations
of motion can be cast in $O(d,d)$ covariant form since the equations of motion
are conservation laws in the superspace.\\
We  argue that the string coordinates and backgrounds, parametrizing target
space $AdS_3$ and $S^3$, transform trivially under the T-duality group
associated with compact directions.Therefore, those equations of motion are
$O(d,d)$ invariant. This completes our study of T-duality symmetry for type
IIB string on $AdS_3\otimes S^3 \otimes T^4$.

\bigskip

\noindent {\bf 4.3 Vertex Operators for Excited States and T-duality}

\bigskip

\noindent We study construction of duality symmetric vertex operators of NSR
string in this subsection.
We have
shown, in the preceding section,
 that for bosonic closed string compactified on $T^d$ the vertex operators
associated with excited massive states can be expressed in an $O(d,d)$
invariant form. This was achieved in a simple frame work. We worked in the weak
field approximation when strings is considered in the background of massive
excited states. These vertex operators were
first expressed in terms  $\sigma$-derivatives of $X^{\mu}$  and
the  canonical conjugate momenta of the compact coordinates and $\sigma$
and/or $\tau$ derivatives. The vertex operators are required to fulfill
following conditions. (i) All vertex operators are required
to be $(1,1)$ operators with respect to the stress energy momentum tensors,
$(T_{++}, T_{--})$, of the free string \cite{ov1,ov1a}.
This is a powerful constraint
and it leads to the 'equations of motion' and 'gauge conditions'
for the massive backgrounds when they are arbitrary functions of
string coordinates. (ii) At each mass level one constructs
'vertex functions' from the basic building blocks such as
$\partial X^{\hat\mu},
{\bar \partial}X^{\hat\mu}$ and $\partial ~{\rm or}~ {\bar \partial}$ acting
on these building blocks. Note that we do not admits terms like
$\partial{\bar\partial}X^{\hat\mu}~ {\rm or}~
{\bar\partial}\partial X^{\hat\mu}$
in vertex functions
since these objects and derivative of such objects vanish as a virtue of
free string equations of motion. (iii) The structure of vertex function of a
given type, at each mass level, is constrained by the level matching conditions
since there is no preferred point on a closed string loop. (iv) The vertex
operator of a given mass level is sum of all such vertex functions. The
vertex operator is required to be $(1,1)$; consequently, at a given mass level,
 the vertex functions are related to each other (see \cite{maha} for details).
At each mass level there are excitations of various
angular momenta of same mass.
In other words the states belong to the irreducible representations of
$SO({\hat D}-1)$. (v) When we compactify the theory to lower dimensions:
$M_D\otimes T^d$, all the states of a given level are classified according to
irreducible representations of $SO(D-1)$ including the states coming from
excitations in compact directions (these are all scalars) since total degrees
of freedom (at each level) remains the same in both the cases.\\

We mention in passing that the constraints of conformal invariance need not be
imposed at this stage while we are investigating duality symmetries. Those
requirements further restrict the structures of of vertex operators and
provide useful relations among vertex functions besides imposing mass shell
conditions for a given mass level.
\\
We intend to derive analogous results for the vertex functions of the excited
massive states of NSR string. Notice that for the first excited level on
the leading Regge trajectory for NSR string will have a lot
more terms compared
to (\ref{firstex}) since we can construct additional terms which contract
with chiral worldsheet fermions. For example, we can have generic terms like
\bea
\label{superf}
G^{(1)}_{\hat\mu\hat\nu\hat\rho\hat\lambda\hat\delta}\partial X^{\hat\mu}
\partial X^{\hat\nu}\psi^{\rho}_-\psi^{\hat\lambda}_-
{\bar\partial}X^{\hat\delta}
,~ G^{(2)}_{\hat\mu\hat\nu\hat\rho\hat\lambda\hat\delta\hat\epsilon}
\partial X^{\hat\mu}\partial X^{\hat\nu}\psi^{\rho}_-\psi^{\hat\lambda}_-
\psi^{\delta}_-\psi^{\hat\epsilon}_-
\eea
and several other terms where $\partial$ is replaced by ${\bar\partial}$ and
$\psi_-$ is replaced by $\psi_+$ so long as we ensure, to start with, we have
maintained same dimensionality for product of left movers and right movers with
respect to the two stress energy momentum tensors. The discrete $Z_2$ symmetry
alluded
to in (\ref{z2duality}) will be maintained if we take into account all required
terms for the vertex operator under considerations. It is quite obvious even
keeping track of all terms for the vertex operators of some of the low lying
excited
massive levels is going to be not very efficient if we want to check
the conjectured T-duality for superstrings in terms of the bosonic coordinates
and NSR fermions. So far there was no construction of manifestly $O(d,d)$
invariant vertex operators for NSR string along compact directions even for
the massless sector i.e. massless scalars that arise from compactification
of $ G_{\hat\mu\hat\nu}$ and $ B_{\hat\mu\hat\nu}$.\\
Therefore, we resort to the superfield approach and consider vertex functions
for massive excited states constructed out of the superderivatives of
superfields \cite{mahans}. There are two types of generic vertex functions
\\
(i) ${\overline D}W^{i_1}{\overline D}W^{i_2}... {\overline D}W^{i_m}
DW^{j_1}DW^{j_2}...
 DW^{j_m}$.These
correspond to leading Regge trajectories.\\
(ii) ${\overline D}^pW^{i_1} {\overline D}^qW^{i_2}....{\overline D}^r
W^{i_m}{D}^{p'}W^{j_1}
{D}^{q'}W^{j_2}...{D}^{r'}W^{j_m}$ and we require $p+q+r=p'+q'+r'$. \\
We proceed with following step for the case first case i.e. vertex functions
corresponding to leading Regge trajectories which are obtained after the
compactification.
(I) Recall that $\bf U$ is an $O(d,d)$ vector, whose upper component is
 $W^i$ and
the lower component is its dual ${\widetilde W}_i$. Introduce two projection
operators \cite{mahans}
\bea
\label{uproject}
{\widetilde P}_+=\pmatrix{1 & 0\cr 0 & 0\cr} ~~~
{\widetilde P}_-=\pmatrix{0 & 0\cr 0
& 1 \cr}
\eea
Note that ${\widetilde P}_+{\bf U}=W$. \\
(II) Introduce a doublet through the pair
$(D,{\overline D})$.
\be
\label{ddoublet}
 {\cal D}=\pmatrix{D \cr {\overline D} }
\ee
Then  projection operators
\bea
\label{projectD}
{\widetilde \Delta}_+=\pmatrix{1 & 0 \cr 0 & 0 \cr} ~~{\widetilde \Delta}_-=
\pmatrix{0 & 0 \cr 0 & 1 \cr}
\eea
Note that ${\widetilde P}_{\pm}$ are $2d\times 2d$
dimensional projectors whereas
${\widetilde \Delta}_{\pm}$ are $2\times 2$ projectors.\\
The vertex functions which assume the form given in (i) above can be cast as
products of $O(d,d)$ vectors \cite{mahans}
\bea
\label{oddtensor1}
{\widetilde \Delta}_-{\cal D}{\widetilde P}_+{\bf U}^{\alpha_1}....
{\widetilde \Delta}_-{\cal D}{\widetilde P}_+{\bf U}^{\alpha_m}
{\widetilde \Delta}_+{\cal D}{\widetilde P}_+{\bf U}^{\beta_1}....
{\widetilde \Delta}_+{\cal D}{\widetilde P}_+{\bf U}^{\beta_m}
\eea
Thus we have an $O(d,d)$ tensor of rank $2m$. We contract it with
a tensor, which depends on spacetime superfield $\phi^{\mu}$ to construct an
$O(d,d)$ invariant vertex function for the $n^{th}$ massive level.
\bea
\label{leadingv}
V_{n+1}=T_{\alpha_1..\alpha_m\beta_1..\beta_m}
{\widetilde \Delta}_-{\cal D}{\widetilde P}_+{\bf U}^{\alpha_1}....
{\widetilde \Delta}_-{\cal D}{\widetilde P}_+{\bf U}^{\alpha_m}
{\widetilde \Delta}_+{\cal D}{\widetilde P}_+{\bf U}^{\beta_1}....
{\widetilde \Delta}_+{\cal D}{\widetilde P}_+{\bf U}^{\beta_m}
\eea
If the $O(d,d)$ vector transforms as:
$U^{\alpha_1}\rightarrow \Omega^{\alpha_1}_{\alpha '_1}U^{\alpha'_1}$, then we
require
\bea
\label{trans}
T_{\alpha_1,..\alpha_m\beta_1..\beta_m}\rightarrow \Omega^{\alpha'_1}_{\alpha_1}
..\Omega^{\alpha'_m}_{\alpha_m}\Omega^{\beta'_1}_{\beta_1}...
\Omega^{\beta'_m}_{\beta_m}T_{\alpha'_1..\alpha'_m\beta'_1...\beta'_m}
\eea
so that the vertex function $V_{n+1}$ is T-duality invariant.\\
Now we focus attention on the second type of vertex function mentioned in (ii)
above. Note that a typical term appearing in the product is like
$({\overline D})^p$ and $(D)^{p'}$.
We can use the projection operators introduced
in (I) and (II) above to express products of such terms as \cite{mahans}
\bea
\label{odds2}
{\widetilde\Delta}_-^p{\widetilde P}_+U^{\alpha_1}{\widetilde\Delta}_-^q
{\widetilde P}_+
U^{\alpha_2}..{\widetilde\Delta}_-^r{\widetilde P}_+U^{\alpha_m}
{\widetilde\Delta }_+^{p'}{\widetilde P}_+U^{\beta_1}
{\widetilde\Delta }_+^{p'}{\widetilde P}_+U^{\beta_2}..
{\widetilde\Delta }_+^{p'}{\widetilde P}_+U^{\beta_m}
\eea
Now this is an $O(d,d)$ tensor of rank $p+q+r$ satisfying the level matching
condition. As in the previous case, we have to just contract with a tensor
(which depends on superfield $\phi$) to get an $O(d,d)$ invariant vertex
function.\\
So far we have left out two other possibilities which we dwell upon now.
There are two more types of vertex functions in a given level: \\
(a) We can have a situation that the vertex function has product of mixed
set of operators i.e. some of the superfields correspond to spacetime
coordinates and some to compact ones.
\bea
\label{mixedvertex}
T_{\mu_1..\mu_k\alpha_1..\alpha_l\mu'_1..\mu'_k\alpha'_1\alpha'_l}(\phi)
{\overline D}^p\phi^{\mu_1}..{\overline D}^q\phi^{\mu_k}{\overline D}^{r}
W^{\alpha_1}{\overline D}^s
W^{\alpha_l}D^{p'}\phi^{\mu'_1}D^{q'}\phi^{\mu'_k}D^{r'}W^{\alpha'_1}D^{s'}
W^{\alpha'_l}
\eea
First notice that $\phi^{\mu}$ is inert under T-duality transformations.
Similarly, all the spacetime indices of the tensor
$T_{\mu\nu...\mu'\nu'..}(\phi)$
do not get transformed under T-duality. Moreover, all the spacetime indices
of $T$ are contracted with spacetime superfields so that effectively we deal
with a tensor with "internal" indices which are contracted with product of
building blocks consisting of superderivatives of $W$'s i.e ${\overline D}~
{\rm or}~ D$
acting on $W$'s . We already presented a prescription of constructing $O(d,d)$
invariant vertex functions out of such products. Therefore, any arbitrary
vertex function of a given massive level can be expressed in an $O(d,d)$
invariant form.
\\
(b) There is another class of vertex functions which are product of the
superderivatives of the spacetime superfields only. However, this class of
vertex functions are automatically T-duality invariant since the
superderivatives $\phi^{\mu}$ and corresponding $\phi$-dependent tensors are
not sensitive to $O(d,d)$ transformations.\\
In conclusion for an NSR string compactified  on $T^d$, we can express all
vertex functions at each massive level in T-duality invariant form.
\\
Another important point deserves attention. Let us consider the the two generic
vertex functions represented by (\ref{leadingv}) and (\ref{mixedvertex}).
A vertex operator at a given mass level is linear combination of all possible
vertex functions consistent with the mass for that level and satisfying the
level matching condition. When we demand that, thus constructed vertex
operator satisfies $(1,1)$ condition with respect to the stress energy momentum
tensor, all the vertex functions are not necessarily independent. The vertex
functions satisfy 'equations of motion' which are onshell conditions and
satisfy 'transversality' conditions. For a given mass level the tensors given
in (\ref{mixedvertex}),
$T_{\mu_1,\mu_2..\alpha_1..\alpha_l,\mu'_1,\mu'_2..\alpha'_1..\alpha'_l}$,
 for example, depends on the superfield, $\phi^{\mu}$, which are along the
noncompact directions. These tensors satisfy 'equations of motion' and the
transverality condition when we demand that the vertex operators satisfy
requirements of superconformal symmetry. Note that of   the set of tensors
appearing in the construction of vertex functions not all are independent. For
detail discussions, in the context  of excited levels of closed bosonic string
we refer the reader to \cite{maha} where T-duality properties were discussed
and the equations of motion and transversality conditions were presented. If
we were to envisage above issues from the perspective of BRST formalism,
the vertex operators will be required to be BRST invariant. As is well known,
both the equations of motion and transversality conditions of the vertex
operators will be derived as a consequence. We mention in passing that this
investigation is focused on NS-NS sector of the theory. It will be interesting
to study the RR sector in this approach. This will be taken up in a separate
paper. \\
We address another point in the context of type IIB theory.
It is well known that this theory
 in endowed with S-duality symmetry. Its massless
spectrum in NS-NS sector for
critical dimension (${\hat D}=10$) consists of graviton,
${\hat g}_{\hat\mu\hat\nu}$, 2-form antisymmetric tensor,
${\hat B}^{(1)}_{\hat\mu\hat\nu}$ and dilaton, ${\hat{\cal\phi}}$. The R-R
sector is axion, ${\hat{\cal \chi}}$, 2-form antisymmetric tensor,
${\hat B}^{(2)}_
{\hat\mu\hat\nu}$ and a four form tensor
${\hat C}^{(4)}_{\hat\mu\hat\nu\hat\rho\hat\lambda}$ whose field strength
is required to be self dual. The effective action for the type IIB theory
may be expressed in an S-duality invariant form. When we toroidally
compactify the theory to lower dimension, the reduced effective can also
written in S-duality invariant form. Therefore, starting from NS-NS
backgrounds, we can generate RR backgrounds of the reduced theory; however,
the reduced tensors of four form $C^{(4)}$ cannot be generated from
the NS-NS backgrounds. Let us closely examine
 the case
when type IIB theory is compactified on $T^4$ to a six dimensional
theory and focus our attention on the moduli and the vector fields
coming from reduction of backgrounds
\footnote{ I am thankful to John Schwarz for elucidating the arguments
presented here} .
 We expect that the
massless states coming from NS-NS sector will be classified according to
representations of $O(4,4)$. In fact the moduli parametrizes the coset
${{O(4,4)}\over{O(4)\otimes O(4)}}$ as  was demonstrated by us \cite{ms}.
The counting is quite simple: the moduli coming from compactification of the
graviton and the 2-form antisymmetric tensor  add up to 16 as
expected. The gauge fields originating from metric and antisymmetric tensor
(from NS-NS sector) belong to the vector representation of $O(4,4)$. Indeed,
in the NS-NS sector the worldsheet action exhibits presence of all these
massless fields, if we follow prescriptions of ref \cite{ms}. Let us turn to
the R-R sector. There are 9 scalar appearing due to compactification of the
2-form, $B^{(2)}$, the 4-form $C^{(4)}$ besides the axion. The number of vector
fields are eight: four from $B^{(2)}$ and four from $C^{(4)}$. We should also
take into account the underlying S-duality symmetry: ${SL(2,Z)}$.
It is more appropriate to classify massless states of the toroidally
compactified six dimensional theory combining the states
from NS-NS and RR sector. The
arguments are along the same line as classification of branes (hence
classifying the background tensors) in the
context of toroidal compactification of type IIB theory and
M-theory \cite{jhs1,jhs2,jm2}.
 They belong to representations of $O(5,5)$ from this
perspective. Whereas the 25 (=16+9) moduli parametrize the coset
${O(5,5)}\over{O(5)\otimes O(5)}$, the 16 (=8+8) massless vectors belong
to the spinor representation of $O(5,5)$.The other backgrounds, in the
six dimensional theory,  also belong
to appropriate representation of this group.\\
 We note that one can study the T-duality attributes of the NS-NS massless
backgrounds of the theory compactified on $T^4$ in the worldsheet approach
presented here. The massive excited backgrounds along the compact direction,
in the NS-NS sector, can be coupled to corresponding worldsheet
supercoordinates. We are able to express the vertex operators for each of
such levels in a manifestly  T-duality invariant form. However, it is not
possible to construct similar vertex operators for the RR sector in the
present formulation.



\section{Summary and Conclusions}

\noindent
We have studied T-duality for compactified closed strings from the worldsheet
point of view. It is argued that this approach reveals some of the salient
features of the symmetry. We have presented explicit examples to demonstrate
some of the interesting properties of this symmetry. Moreover, we reviewed
the role the K-K modes and winding modes when the string is compactified
on $T^d$ in the presence of constant $G_{\alpha\beta}$ and $B_{\alpha\beta}$.
We demonstrated that the spectrum remains invariant under $O(d,d)$
transformations and showed how the discrete $O(d,d, Z)$ symmetry appears
due to periodicity of the compactified closed string coordinates on $T^d$. Next
we reviewed the duality symmetry of string evolution equations on the
worldsheet. We considered the scenario where, the string coordinate
$X^{\hat\mu}$ are decomposed to sum of noncompact spacetime coordinates,
$X^{\mu}, \mu=0,1...D-1$ and compact coordinates $Y^{\alpha}, \alpha=D, D+1,...
{\hat D}-1$ so that $D+d={\hat D}$. Moreover, all backgrounds are independent
of $Y^{\alpha}$ and depend only on $X^{\mu}$. We showed that the equations of
motion associated with are $O(d,d)$ covariant and those associated with
$X^{\mu}$ are $O(d,d)$ invariant. This sets up a background to provide a better
understanding of our approach for the study of duality symmetry associated
with excited massive levels.\\
We have proposed a systematic procedure to obtain T-duality invariant
vertex functions for massive levels of a closed bosonic string when it is
compactified on $T^d$, the $d$-dimensional torus. It is assumed that the
tensor fields associated with these vertex operators depend only on the
spacetime coordinates, $X^{\mu}(\sigma,\tau)$ and are independent of the compact
coordinates, $Y^{\alpha}(\sigma,\tau)$. The duality invariance is manifest for
vertex operators of each level once one uses the projection technique to
convert $\{P,Y'\}$ to $O(d,d)$ vectors and/or their $\Delta_{\pm}$ derivatives.
However, this programme is not complete. We have considered the Hassan-Sen
compactification \cite{hs} where the metric is decomposed into two diagonal
blocks. If we adopt the compactification of \cite{ms} then the gauge fields
associated with the isometries have to accounted for. Moreover, we considered
the target space metric to be flat and have set 2-form tensor to zero in the
massless sector. Furthermore, we have ignored the presence of winding modes.
A complete study should incorporate all these factors in investigating
T-duality properties of excited massive level. If history is a guide, we can
conjecture that the vertex operators will continue to be T-duality
invariant in the presence all the background field which we have accounted for
in the present case. We may recall that simple cases in cosmological scenario
were useful in unraveling the $O(d,d)$ symmetry both in the worldsheet
approach and in study of (cosmological) effective action. Moreover,
Hassan-Sen \cite{hs} was an early step to reveal $O(d,d)$ symmetry before more
general programme was undertaken \cite{ms}.  \\
The T-duality symmetry plays an important role in string theory. We expect that
these symmetry properties will have important applications. Recall that
the T-duality symmetry has been widely applied to obtain new solutions to the
background configurations through judicious implementations of the solution
generating techniques. Thus given a configuration of  massive level
background field it will be possible, in principle,  to generate another
background within the same massive level. Furthermore, there are evidences
that massive excited states are endowed with local symmetries. It is worth
while to examine the implications of T-duality for those local symmetries.
\\
Another  point which deserves attention is to study the zero-norm
states in this formulation.
It is well known that the existence of zero-norm states is quite essential
in order that the bosonic string respects Lorentz invariance in critical
dimensions i.e. ${\hat D}=26$. This issue has been carefully analyzed in
\cite{ov1,ov1a}. We expect that these results will continue to hold good
when we are dealing with toroidally compactified closed bosonic string.\\
It is well known that very massive stringy states have possess exponential
degeneracy which has played crucial in deriving Bekenstein-Hawking entropy
relation for stringy back holes from the counting of microscopic states.
This high degree of degeneracy is also instrumental in deducing the
thermal nature of emission spectrum of a stringy black hole. We expect that
some of supermassive states which also belong to the spectrum of the
compactfied string might exhibit symmetry properties which are yet to be
discovered.\\
It is worth while to dwell on another aspect of $O(d,d)$ invariant form of
the vertex operator. We recall that a generic vertex function is expressed
as product of $O(d,d)$ vectors ${\cal W}$ and this product is contracted
with a tensor so that the resulting vertex function is $O(d,d)$
invariant. Note, however, that the the product of the ${\cal W}$-vectors
can be expressed as sum of tensors which are irreducible representations of
$O(d,d)$. Thus the vertex function will be sum of terms which are $O(d,d)$
invariant on their own. For sake of definiteness focus on the vertex function
associated with the first massive level. This serves as an illustrative
example
$A^{(1)}_{\alpha\beta,\alpha'\beta'}\partial Y^{\alpha}\partial Y^{\beta}
{\bar\partial} Y^{\alpha'}{\bar\partial} Y^{\beta'}$. We can always rewrite it
as
\bea
\label{irr}
T^{(1)}_{kl,k'l'}(X){\cal W}_+^k{\cal W}_+^l{\cal W}_-^{k'}{\cal W}_-^{l'}
\eea
We have distinguished the appearance of the $O(d,d)$ vectors in the
expressions for the vertex operators whether they originate
 from right moving or left moving sector through unprimed and primed
indices. The vertex operator (\ref{irr}) is $O(d,d)$ invariant; however,
it could be decomposed as sum of contracted tensors belonging to
irreducible representations of $O(d,d)$. We first illustrate the point
by a simple example from atomic/nuclear physics when one considers
familiar multipole operators which usually appear in computations of
radiative transitions. The operator ${\bf{x^ix^j}}$ is decomposed into
\bea
\label{quadro}
{\bf x^ix^j}=({\bf{x^ix^j}}-{1\over 3}{\bf \delta^{ij}}{\bf x}^2)+
{1\over 3}{\bf x}^2{\bf{\delta^{ij}}}
\eea
Note that the first term is the quadrupole operator (traceless). If we
construct  a scalar ${\bf T_{ij}x^ix^j}$; the product decomposes into
${\bf{ T_{ij}Q^{ij}}}+{\bf T_i^i}{\bf x}^2$; ${\bf Q^{ij}}$ being the
quadrupole operator.\\
Let us examine the tensor structures in (\ref{irr}). $T^{(1)}_{kl,k'l'}$
is contracted with product of ${\cal W}_+^k{\cal W}_+^l$ and
 ${\cal W}_-^{k'}{\cal W}_-^{l'}$. Each of these tensors can be
decomposed as follows
\bea
\label{irrdd1}
{\cal W}_+^k{\cal W}_+^l &=&\bigg({\cal W}_+^k{\cal W}_+^l-{1\over{2d}}
{\bf \eta}^{kl}{\cal W}_+^m{\bf\eta}_{mn}{\cal W}_+^l\bigg)+
{1\over 2d}{\cal W}_+^m{\bf\eta}_{mn}{\cal W}_+^l
\eea
and
\bea
\label{irrdd2}
{\cal W}_-^{k'}{\cal W}_-^{l'}=\bigg({\cal W}_-^{k'}{\cal W}_-^{l'}
-{1\over{2d}}{\bf \eta}^{k'l'}{\cal W}_-^{m'}{\bf\eta}_{m'n'}{\cal W}_-^{l'}
\bigg)+{\cal W}_-^{m'}{\bf\eta}_{m'n'}{\cal W}_-^{l'}
\eea
It is obvious that (\ref{irr}) will be composed of sum of terms arising from
IRR of $O(d,d)$,
taking into account
the decompositions (\ref{irrdd1}) and (\ref{irrdd2}),  which are generalization
of (\ref{quadro}) for the case at hand.  Therefore, a generic vertex operator,
for the
$n^{th}$ massive level, which assumes the
form (\ref{generic}), can be converted to an expression of the type
 (\ref{vertexn}) using our prescription. Since they are eventually expressed
as products of $O(d,d)$ vectors and contracted with suitable tensors.
These product of the the $O(d,d)$ vectors will be decomposed into direct
sums of the IRR of $O(d,d)$ and thus will contract with the decomposed
$O(d,d)$ tensors written as direct sums of IRR tensors.
We conclude that all the vertex operators of each massive
level will be expressed as sums of IRR's of the T-duality group. Thus as we
go to higher and higher levels, we have to deal with higher and higher
dimensional representations of this noncompact group.
\\
In Section (IV) we focused attention on superstring in the NSR
superfield formulation.
We presented two results. We  considered NSR string in its massless excitations
such as graviton and antisymmetric tensor coming from the NS-NS sector.
Here we adopted Hassan-Sen compactification and assumed that the backgrounds
along compact directions  are independent of superfields along compact
directions. The equations of motion corresponding to compact superfields can
be cast in $O(d,d)$ covariant form once we introduce dual superfields
corresponding to compact superfields and define corresponding dual
backgrounds. Therefore, we now have derived duality covariant worldsheet
equations of motion which is analog of the equations for closed bosonic
string.\\
This step was very useful to construct vertex operators in the NS-NS background
for excited massive states. We introduced projection operators to discuss
duality properties of vertex functions of massive levels. Therefore,
for the Hassan-Sen compactification scheme, we were able to cast the vertex
functions in $O(d,d)$ invariant form. However, this programme is not complete
so far. We have to address issues in the R-R sector. Next, in order to
construct vertex functions for superstrings (type IIA and type IIB), we have
to analyze role of GSO projection carefully. An important issue deserves
mention here. If we are to carry forward this programme to superstrings, then
it is more appropriate to adopt BRST prescription for vertex operators.
As is well known, in this formalism, one can a suitable picture. Now the vertex
operator is required to be BRST invariant. It is quite likely, when the
vertex operators are constructed for superstring, BRST formulation will be
economical and more elegant.\\
We presented an example where our proposal for NSR string can be concretely
realized. We considered type IIB compactified on $AdS_3\otimes S^3\otimes T^4$.
Here $AdS_3$ is the spacetime. This simple compactification has certain
advantages for us. First of all we can have NS-NS 3-form flux along $S^3$. We
can have constant NS-NS flux along $AdS_3$ with opposite strength. Therefore,
in the worldsheet formulation (superfields), we note that the
$\sigma$-model associated with $AdS_3$ and $S^3$ are WZW models on group
manifolds $SL(2,R)$ and $SU(2)$ respectively and WZ term accounts for constant
$H=dB$ on these manifolds. Moreover, the conformal point the conservation
laws in each case are conservation of holomorphic and antiholomorphic super
current. Furthermore, we can write down the worldsheet action for string along
$T^4$. Note that the target manifold is a direct product. Therefore, we can
write down the vertex operators for excited massive levels following our
prescriptions. Although we have focus attention of Ns-NS sector, an
S-duality transformation will take us from NS-NS field strengths to the R-R
field strength. We mention in passing that all the branes, for this
compactification, belong to representation of the $U$-duality group
$O(5,5)$. Indeed the coset structure is ${O(5,5)}\over{O(5)\otimes O(5)}$
 However, we have no insight to present worldsheet realization of this symmetry. \\
There is another interesting approach to dualities in the worldsheet
approach. In this formulation the number of string coordinates are doubled and
this in scenario some of the nice features of conventional worldsheet approach
are not maintained; however, it has been argued that such doubling might
have deep significance
\cite{duff,witten,wittena,wittenb,wittenc} in string theory. This approach has
not made much headway in string theory, the mathematics is indisputable.
In this background,
 recently, a new formulation
of  field theory has been introduced where $O(D,D)$ invariant action is
constructed, $D$ being the number of spacetime dimensions which is doubled
\cite{double1,double1a,double1b,double1c,double2}. A lot of progress is being
made in double field theory with very beautiful mathematical structures.
At this stage we have not been able to establish
connection of our worldsheet formulation and $O(d,d)$ symmetry
with double field theory.

\bigskip

\noindent
{\bf Acknowledgments:} I have benefited from discussions with members of 
String Theory group at Institute of Physics and at NISER; especially from 
Anirban Basu and Yogesh Srivastava.  I would like to thank Caltech and CERN 
Theoretical Physics groups for providing stimulating atmosphere 
during various stages of this work and for their
very gracious and warm hospitality. I acknowledge very valuable discussions
with John Schwarz and Gabriele Veneziano and thank them for sharing their deep
insights. I would like to thank the String Community of India for their supports
during very trying period.  This work is supported by the People of the Republic
of India through a Raja Ramanna Fellowship of DAE.

\end{document}